\documentclass[12pt]{iopart}
\usepackage[lmargin=1.0in,rmargin=1.0in,tmargin=.8in,bmargin=.8in]{geometry}
\usepackage{graphicx}
\usepackage{xcolor}
\usepackage[utf8]{inputenc}
\usepackage{indentfirst}
\usepackage{longtable}
\usepackage{iopams}
\usepackage{enumerate}
\DeclareUnicodeCharacter{2009}{\,}
\DeclareUnicodeCharacter{2009}{\,}
\usepackage[columnwise,modulo]{lineno}
\usepackage{multirow}
\usepackage{subcaption}
\usepackage[numbers, sort, comma, square,compress]{natbib}
\usepackage{color, colortbl}
\usepackage{subcaption}
\usepackage{url}
\usepackage{booktabs}
\usepackage{enumitem}
\usepackage{varioref}
\usepackage{arydshln}
\usepackage{hyperref}
\hypersetup{colorlinks=true, pdfstartview=FitV, linkcolor=false, citecolor=blue, plainpages=false, pdfpagelabels=false, urlcolor=blue,allcolors=blue}
\usepackage[all]{hypcap}
\newlist{arab}{enumerate}{10}
\setlist[arab]{label*=\arabic*.}

\usepackage{arydshln}

\begin{document}

\title[Electrical charge decay on dielectric surface]{Electrical charge decay on dielectric surface in nitrogen/C$_4$F$_7$N mixtures}

\author{D. Prokop$^{1,2,\ast}$, M. Mrkvi\v ckov\'a$^{1}$, J. Tungli$^{3}$, Z. Bonaventura$^{1}$, P. Dvo\v r\'ak$^{1}$, S. Kadlec$^{3}$ and T. Hoder$^{1,\ddagger}$}

\ead{$^{\ast}$david.prokop@mail.muni.cz, $^{\ddagger}$hoder@physics.muni.cz}

\address{$^{1}$Department of Plasma Physics and Technology, Faculty of Science, Masaryk University,
	Kotl\'a\v{r}sk\'a 2, 611 37 Brno, Czech Republic}

    \address{$^{2}$ELI Beamlines Facility, The Extreme Light Infrastructure ERIC, Za Radnic\'i 835, 25241 Doln\'i B\v re\v zany, Czech Republic}
	
	\address{$^{3}$Eaton European Innovation Center, Bo\v rivojova 2380, 252 63 Roztoky, Czech Republic}
	
\vspace{10pt}

\begin{abstract}
\noindent 

The decay of electrical charge on a dielectric surface in nitrogen/C$_4$F$_7$N (Novec 4710, C4) mixtures is investigated using measurement of electric field via in-situ electric field-induced second harmonic (EFISH) technique. 
The charge is deposited on the surface of the alumina by generating a barrier discharge in the gap, and the amount of charge is determined from electrical current measurements and numerical modeling. 
For different admixtures (0\%, 10\%, and 50\%) of C$_4$F$_7$N in nitrogen, the presence of surface charge is detected even 60 hours after charge deposition. 
It is found that C$_4$F$_7$N admixture lead to a significantly longer-lasting surface charge, indicating a slower charge decay. 
Using an isothermal charge decay model, charge traps are identified for pure nitrogen charge deposition, which are in agreement with results found in the literature. 
Charge deposition in C$_4$F$_7$N admixtures leads to modification or creation of new traps with higher trap energies.
The EFISH measurements are used to determine the C$_4$F$_7$N nonlinear hyperpolarizability tensor component $\alpha^{(3)}_\mathrm{C_4F_7N, \parallel} (-2\omega; 0, \omega, \omega)$. 
Direct comparison of the experimental results from two developed methods (EFISH and electrical measurements) and the numerical model gives a closer insight into the surface charge spread over the dielectrics, resulting in surface charge density estimation.
\end{abstract}

\vspace{2pc}


\maketitle

\section{Introduction}

Electrical charge deposited on a dielectric surface by gas discharges is a phenomenon observed in various contexts, ranging from fundamental studies \cite{Stollenwerk} to applied fields of low-temperature plasma physics. Examples include applications in air flow control \cite{Leonov_2016} and in electrical engineering systems utilizing gas/solid insulation \cite{niem1995}. 
In particular, devices such as medium-voltage switchgear and circuit breakers must withstand prescribed voltage levels without experiencing breakdowns or frequent partial discharges \cite{bartnikas2002}. The threshold voltage of these devices can be influenced by surface charges accumulated as a result of previous  discharge activity.

The motivation for this work is the clarification of the surface charge decay on a dielectric surface in novel insulating gases of acceptable toxicity \cite{rabie2018}. 
Particularly, it is important for the development of ultra-fast, gas-filled,  miniaturized relays capable of carrying high currents ranging from tens to hundreds of amps, designed for use in novel AC and DC hybrid circuit breakers. 
The relay can be filled with various gases - air, nitrogen, SF$_6$, or in mixtures of its alternative, eco-friendly, gases, such as the fluoronitrile C$_4$F$_7$N. 
The relay contacts are fully open to a gap of the order of 0.1\,mm after a given semiconductor bypass time (of the order of 0.1 ms) \cite{horky2024Holm}. 
The dielectric strength of the filling gas is of high importance, as well as the effects of the electrical charge deposited on nearby dielectric surfaces. 
The fluoronitrile C$_4$F$_7$N  (also known as C4 gas or under the trade name Novec 4710) has been studied as a possible alternative to SF$_6$ due to its high dielectric strength and lower global warming potential \cite{rabie2018,vemul2023,liu2025}.

It has been shown that residual surface charge can remain on the dielectric surface for many hours, days or even weeks \cite{navarro2023,Brandenburg}. 
The charge may also redistribute itself laterally on the surface if it is not homogeneously deposited or decay under light irradiation \cite{wild2014,tschiersch2014}. 
Different models of the surface charge decay were developed and their sensitivity to surface and volume conductivities defining the system was analyzed \cite{Emelyanov_model}. 
Taking a microphysical point of view, charge trapping states were also investigated \cite{svandova2024,zhang2018} for alumina dielectrics. 
The surface charge decay is a complex phenomenon including multiple decay processes. Different models describing it macro or microscopically can be applied under various conditions \cite{kinder2008,molinie2024}.

The phenomena of electrical charge deposition on dielectric surfaces, its quantification and decay, has been investigated in detail by various experimental methods. 
The charge transfer to surface via an electrical discharge and the related gas gap voltage may be determined using the electrical current measurements combined with equivalent electrical circuits \cite{tschiersch2014,mrkvi2023}, yet this is usually applied for periodically occurring discharges and not for sensing the charge decay or changes in the residual gap voltage long after the last discharge is gone. 

The utilization of electrostatic probes has proven to be a powerful technique for quantifying the spatial distribution of residual surface charges with high spatial resolution, particularly when enhanced by additional signal processing \cite{Kumada}. This method also enables the determination of the local electric field induced by surface charge domains. 
Importantly, the spatial distribution of the surface charges and its development may enable to identify the charge decay processes \cite{zhang2018}. 
There are possible drawbacks of this method, e.g. reading the signal from the studied dielectric surface may take several minutes, or the sample may need to be transferred from the discharge cell to the electrostatic probe site. 
This is a significant consideration when working with gases that may pose health risks if decomposed in plasma, although this can be solved by placing the discharging electrode and the probe within one vessel. 
Finally, some probes can cause partial distortions in the residual charge distribution or be influenced by electrostatic interaction with environment \cite{kinder2008b,llovera2009}.

Other techniques, based on the Pockels effect, enable not only the determination of spatial distribution of the residual charges but also its temporal development \cite{zhu1996}, even on sub-nanosecond scale \cite{kumada2002}. 
It was recently applied in multiple setups and for various applications, see for example \cite{wild2014,tschiersch2014,li2022}, even for simultaneous temperature measurements via Mueller polarimetry \cite{slikboer2018}. 
The disadvantage of this method is the requirement to use specific materials, typically Bi$_{12}$SiO$_{20}$ (BSO), as dielectrics that exhibits the Pockels effect. 
As different materials exhibit different surface charge decay \cite{pribyl2024}, the necessity to use BSO may be limiting factor, eventhough new hybrid approaches are proposed \cite{limburg}. 
Moreover, BSO is a very fragile material and can break easily if not handled with care.

To address the challenges associated with the aforementioned methods, to enable in-situ or in-operando measurement of charge decay over extended periods without requiring special materials and distorting the sample, we employ the electric field-induced second harmonics (EFISH) technique. 
The EFISH technique was recently introduced for its use in low-temperature plasmas by Dogariu et al. \cite{dogariu2017}. 
The method was further developed in various laboratories, several issues and challenges connected with its application were solved, and the method was significantly improved, see \cite{chng2020electric,chng2022effect,nakamura2021,orr2020,raskar2022,billeau2024, mrkvi2023}. 
This laser-aided diagnostics allows for the measurement of the local electric field in close vicinity of the dielectrics, in this case alumina, of the barrier discharge cell. 
Electric field magnitude is directly linked to the presence of residual surface charges on the dielectrics and it enables to determine the residual voltage using a novel compensation method, as we present in this manuscript. 
It is done by changing the external voltage applied on the cell, which compensates the electric field effect of surface charges in the laser-probed coordinate. 
The value of the applied voltage causing the minimal value of resulting EFISH signal is called the compensating voltage, it is the direct measure of the residual surface charge. 
Parallel to this novel approach, another method for residual charge determination is developed and presented - using simple electrical current and voltage measurements. It shows a good agreement with EFISH based method.

In this article, we present experimental results on the quantification of the charge transferred through the gas gap using fast electrical current measurements. The compensating voltage is subsequently determined during both the primary (forward) and the backward single micro-discharges. 
These measurements are conducted in a barrier discharge setup where only one electrode is covered with an alumina dielectric.
The working gas are mixtures of nitrogen with C$_4$F$_7$N gas, particularly 0\%, 10\% and 50\% of C$_4$F$_7$N in pure nitrogen. 
After the backward discharge, the decrease of the compensating voltage magnitude in time due to the residual charge decay (the so called surface charge/potential decay) on dielectric surface is measured and first estimations discussing the physics responsible for that are given using an isothermal charge decay model. 
The presented results enable also determination of the C$_4$F$_7$N gas hyperpolarizability tensor component $\alpha^{(3)}_\mathrm{C_4F_7N, \parallel} (-2\omega; 0, \omega, \omega)$. 
The charge deposition caused by barrier discharge streamers is compared with a two-dimensional (2D) streamer model, allowing the determination of the deposited charge surface density and its effective spread over the dielectric surface. 
Both the EFISH and electrically determined compensating voltages are analyzed theoretically, too.

The presented methodology, which combines experimental and theoretical analyses, is promising in-situ, cross-validated approach for investigation of surface charge decays also in other gas mixtures on various insulating materials for future electro-technical applications.

The paper is organized in the following way. In the next section, the experimental methods, materials and setup are described. In the third section the methodology of the surface charge decay measurements is presented. 
The fourth section introduces the model used for simulations of the streamer discharge responsible for the charge deposition on the surface. 
The fifth section present our results on residual charge decay and discussions. 
In the last section we summarize and conclude the paper.

\begin{figure}[!h]
    \centering
\includegraphics[width=0.9\textwidth]{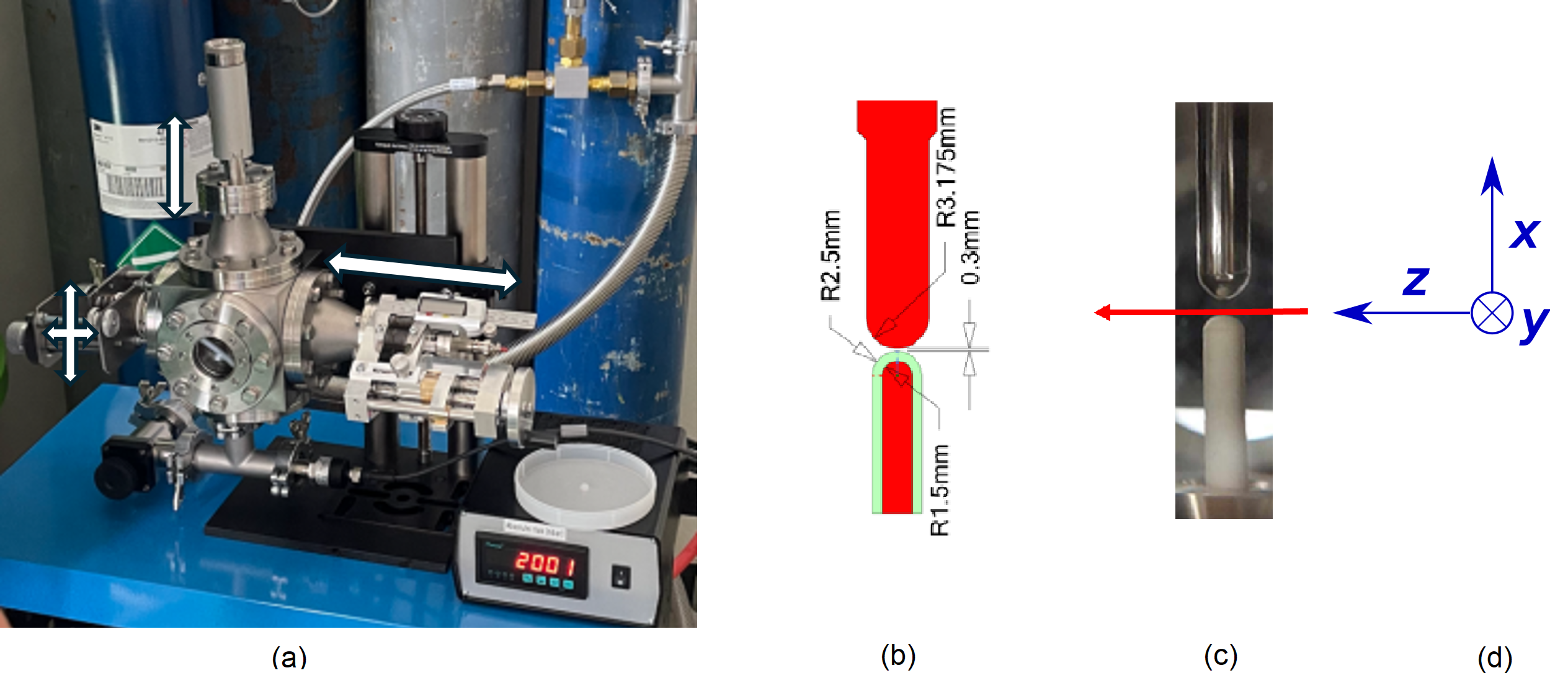}
\caption{Experimental setup, showing the vacuum sealed test cell equipped with stainless steel and Al$_2$O$_3$ covered electrode system with the possibility of setting the gas gap in the order of tens of micrometers, in part a). In part b), the scheme of the electrodes (steel in red, alumina in light green) used with their parameters is shown. In part c), the  positioning of the laser beam in the electrode gap is depicted, 100\,$\mu$m above the dielectric surface. The selected coordinate system is depicted in d) with laser propagating along positive $z$ axis and position in the inter-electrode gap marked as $x$ coordinate (with $x$ = 0 at the dielectric surface). Note that the electrodes in the chamber are positioned horizontally in the laboratory framework and the laser enters the chamber from the front window, as depicted in part a). }
\label{komora}
\end{figure}

\section{Experimental setup}
The experiments were performed in a sealed stainless steel chamber shown in Fig.\,\ref{komora}a). The chamber is designed for testing of electrical properties of various gases and allows operation at both lower and higher than atmospheric pressure (from 2\,mbars up to 2\,bar). 
The chamber has been filled with pure nitrogen (purity of 99.996\%) and with its mixtures with C$_4$F$_7$N: 0\%, 10\% and 50\% of C$_4$F$_7$N in the nitrogen gas. 
After every experimental series, the chamber has been evacuated and refilled. 
All experiments were done at atmospheric pressure and laboratory temperature of approximately 300\,K. 

The chamber operation enables application of high-voltage pulses up to 20\,kV and is equipped with an interchangeable system of two opposing electrodes. The main feature of the chamber is the electrode manipulation system which allows individual positioning of both electrodes with an accuracy of tens of micrometers. 
The chamber features two optical view-ports (fused silica) for imaging, optical emission and laser spectroscopic measurements.

\begin{figure}[h!]
    \centering
    \begin{subfigure}[t]{0.49\textwidth}
        \centering
        \includegraphics[width=\textwidth]{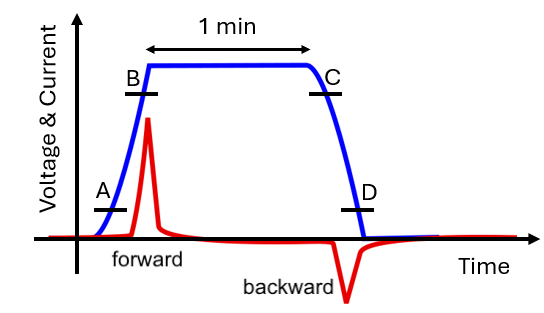} 
        \caption{}
        \label{fig:sub2}
    \end{subfigure}
    \begin{subfigure}[t]{0.49\textwidth}
        \centering
        \includegraphics[width=\textwidth]{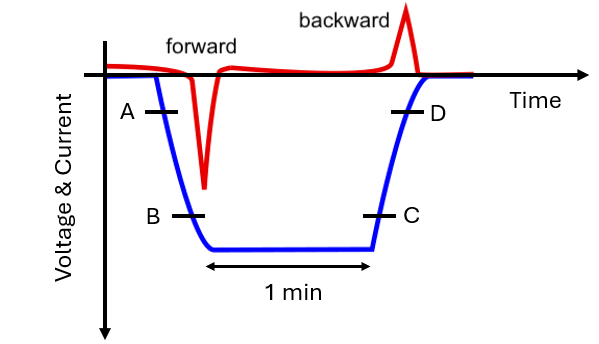} 
        \caption{}
        \label{fig:sub3}
    \end{subfigure}

    \caption{Schematic representation of the voltage (blue) and current (red) waveforms for a) positive polarity, b) negative polarity of the applied voltage. Rise time of the voltage waveform (time between A and B) was about 1 s. Falling time (time between C and D) was about 5 s. The first discharge is denoted as forward and the second as a backward one.}
    \label{fig:scheme_pulse}
\end{figure}

The electrode arrangements consisted of high-voltage hemispherical stainless-steel electrode (radius 3.175\,mm) and grounded  electrode covered by Al$_2$O$_3$ ceramics (inner radius 1.5\,mm, outer one 2.5\,mm, i.e. thickness of 1\,mm, specification C795 or AG202, purity better than 95\,\%), see Fig.\,\ref{komora}b). 
The gap size was adjustable. 
For charge decay measurement the gap size was selected so that dominantly one single discharge (forward discharge) followed by one backward discharge occurred during the on-set and off-set of the high-voltage, see the procedure in Fig.\,\ref{fig:scheme_pulse} and in  the text further. 
The positive applied voltage polarity (see Fig.\,\ref{fig:scheme_pulse}a) means that the metallic high voltage electrode was anode, while the dielectrics-covered electrode was anode. For the other polarity it was the opposite.

For the EFISH measurements the above mentioned procedure resulted in selection of the gap size with value of 0.3\,mm. 
Smaller gap sizes caused multiple breakdowns in both directions, while for larger gaps, there was no backward discharge and the behavior was highly unpredictable or discharge not possible due to the high dielectric strength of C$_4$F$_7$N gas.

\begin{figure}[!h]
    \centering
\includegraphics[width=.9\textwidth]{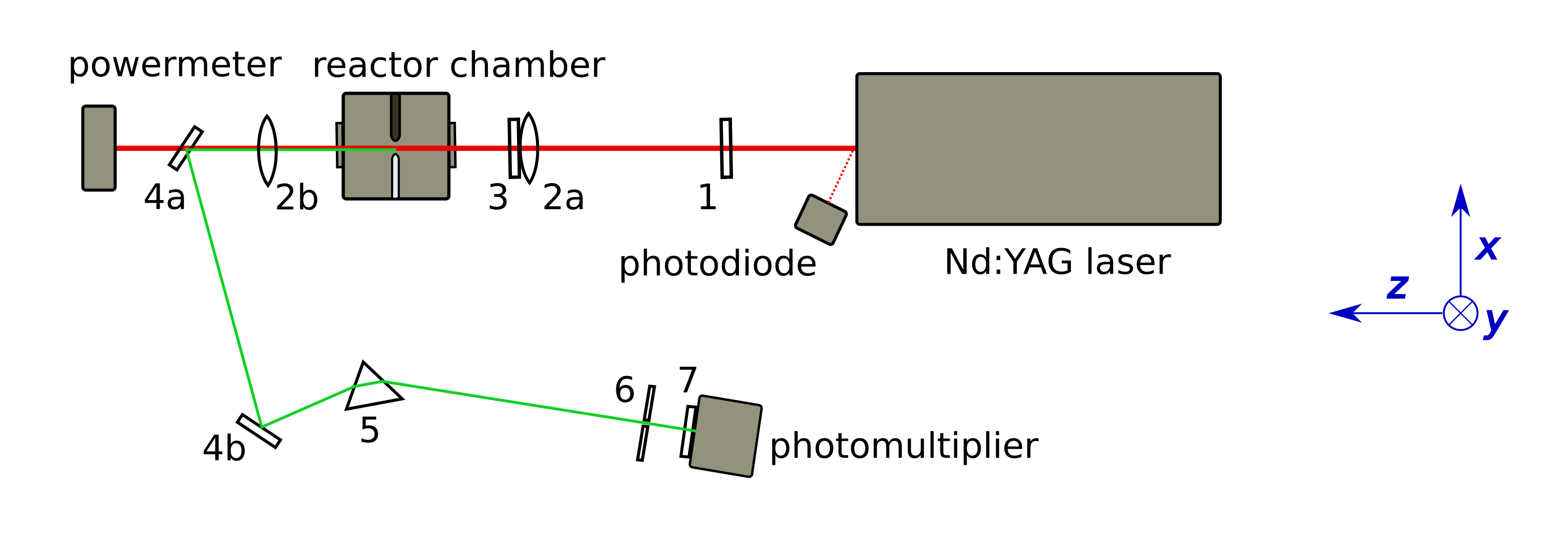}
\caption{Experimental setup for EFISH: 1 -- a half-wave plate, 2 -- two $f=10$\,cm lenses to focus and collimate the laser beam, 3 -- a long-pass filter (700 nm), 4 -- two dichroic mirrors, 5 -- a prism, 6 -- a diaphragm, 7 -- a bandpass filter (532 nm).}
\label{f:efishaparatura}
\end{figure}

As a voltage source we used a DC power supply (FuG Elektronik GmbH, HCP 1400-20000 MOD). The voltage measurements were carried out using a P6015A voltage probe (Tektronix) with maximum input voltage 20\,kV, and bandwidth up to 75\,MHz.
For the current measurements, we have utilized a self-assembled coaxial probe (called BNC probe) measuring the current as a voltage variation on resistor of 50\,$\Omega$, assembled from four 200\,$\Omega$ resistors in star-like connection at the tip of BNC cable. 
The setup and electrical characterization of the probe was already described in \cite{BNC}.

The experimental setup for EFISH measurements is shown in Fig.\,\ref{f:efishaparatura}. As a laser source, we used a diode-pumped high energy picosecond Nd:YAG laser PL2230 from Ekspla. The laser was operated at 1064\,nm with energy less than 0.5\,mJ and repetition frequency of 50\,Hz. Higher energy of laser pulses caused sparks in the discharge gap producing artifacts in the EFISH signal and causing the artificial decay of the surface charge. 
The laser beam passed a half-wave plate to rotate the polarization to horizontal (the $x$ direction is in reality horizontal in the laboratory framework), and it was focused by $f=10$\,cm lens, resulting in a beam waist of 17\,$\mu$m and Rayleigh range $z_R = (8 \pm 4) \cdot 10^{-4}$\,m. The focal point was situated just at the interelectrode axis, 0.10\,mm from the tip of the dielectric surface. 
The position of the laser focus was set with an accuracy of 0.02\,mm along the $x$-axis, 0.5\,mm along the $y$-axis and 1\,mm along the $z$-axis, 
and verified firstly by raising the laser energy above 1\,mJ and observing the position of the spark in the gas, secondly by moving the reactor on the translation stage and observing the laser energy loss due to reflection on the electrodes. 
After exiting the reactor, the beam was collimated by another $f=10$\,cm lens. The IR part of the primary beam was absorbed by a powermeter (Ophir, Vega), whereas the EFISH signal (wavelength of 532\,nm) was filtered out using a pair of dichroic mirrors, a prism with diaphragm and a bandpass interference filter (Thorlabs FL532-10) and has been detected via PMT from Photek (response FWHM to 100\,ps, single photon jitter to 28\,ps). 

All electrical characteristics, whether electrical current and voltage measurements or signals from the laser setup, were captured using a high-definition oscilloscope, Keysight DSO-2 204A (bandwidth 6\,GHz, 10-bit ADC, 20\,GSa/s).

\section{Methodology of the surface charge decay measurements} \label{methodology}

The idea of this work was to study the long-term decay of a surface charge on dielectrics (deposited by the forward and backward discharges resulting from high voltage pulse application) via monitoring of the electric field caused by this charge. The electric field was studied by the method EFISH. When a  laser beam is focused to the investigated electric field area filled with gas, the non-linear response of the media polarisation results in the generation of the second harmonic frequency signal. Following \cite{chng2020electric} and \cite{chng2022effect}, the power of this second harmonic signal for a focused probe beam is given by
\begin{equation}\label{eq_efish}
P^{(2\omega)} \propto \left[ \alpha^{(3)}(-2\omega, 0, \omega, \omega) \cdot N \cdot P_0^{(\omega)} \right]^2 \cdot E_0^2 \cdot z_R \cdot \left| \int_{-\infty}^{\infty} E'_{\mathrm{ext}}(z') \cdot \frac{\mathrm{exp}(i \cdot \Delta k \cdot z_R \cdot z')}{[1 + i \cdot z']} \mathrm{d}z'  \right|^2,
\end{equation}
\begin{equation}
E'_{\mathrm{ext}}(z') = \frac{E_{\mathrm{ext}}(z)}{E_0},
\end{equation}
\begin{equation}
z' = \frac{z}{z_R},
\end{equation} 
where $N$ is the gas number density, $P_0^{(\omega)}$ is the power of the probe beam (1064\,nm),  $z_R$ is the Rayleigh range of the focused laser beam, $E_{\mathrm{ext}}(z)$ is the externally applied electric field distribution along the beam propagation axis $z$, $E_0$ is the electric field strength at $z=0$, $\Delta k = [2k^{(\omega)}-k^{(2\omega)}]$ is the wave-vector mismatch. 
$\alpha^{(3)}$ is the third-order nonlinear hyperpolarizability of the gas, i.e., third-order tensor depending on molecular dipole moments and field orientations \cite{Sitz68, dogariu2017}. 
There are two nonzero components of the hyperpolarizability, $\alpha^{(3)}_\parallel$ corresponding to the laser polarized parallel to the electric field, and $\alpha^{(3)}_\perp$ for the laser polarized perpendicular to the electric field. 
As $\alpha^{(3)}_\perp \approx \frac{1}{3}\alpha^{(3)}_\parallel$ \cite{Irving74}, the parallel component of the electric field will be reflected in EFISH signal with the factor of 9 higher than the perpendicular component. 
In our experiment, the laser beam is polarized horizontally, i.e. along the inter-electrode $x$ axis which is the direction of the dominant component of the electric field. 
Therefore, $\alpha^{(3)}_\perp$ has a negligible effect on the resulting EFISH signal, and we will consider only the parallel component of hyperpolarizability, $\alpha^{(3)}_\parallel$.

\begin{figure}[h]
    \centering
\includegraphics[width=.6\textwidth]{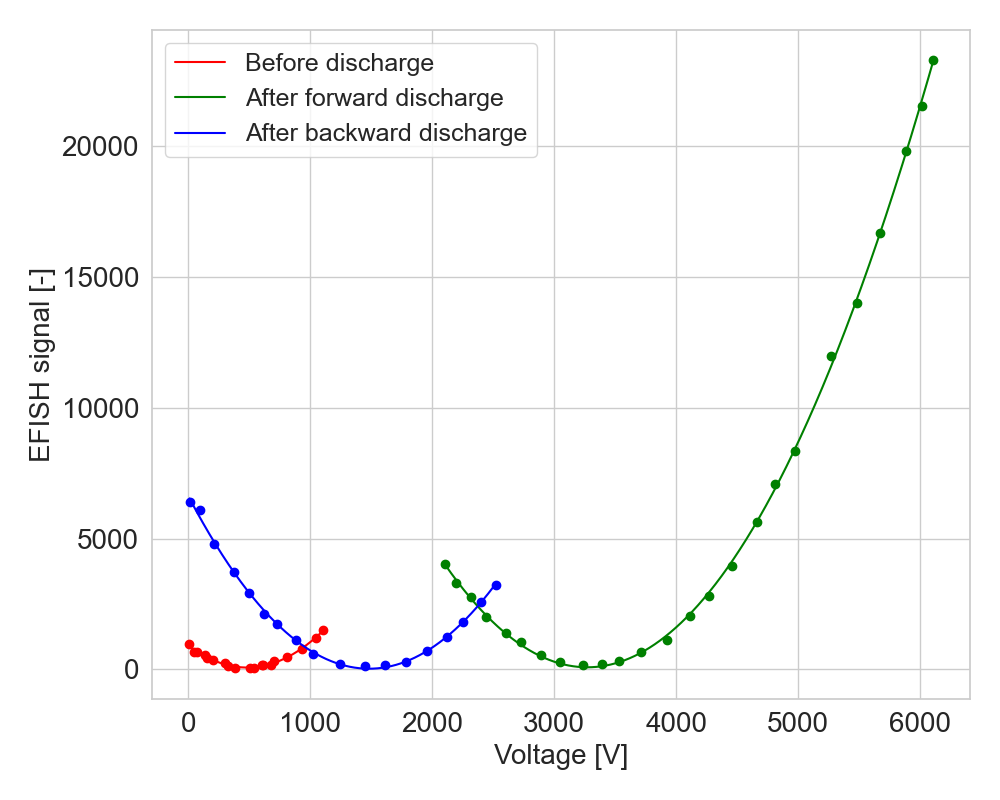}
\caption{EFISH measurement of the signal given by equation\,\ref{eq_S} in pure nitrogen. The signal exhibits quadratic dependence on the applied voltage as described in equation\,\ref{eq:EFISH_quad}.} \label{f_efishparaboly}
\end{figure}

The formula\,\ref{eq_efish} consists of an integral dependent both on the spatial shape of the electric field and on the parameters of the laser beam. The spatial distribution of the surface charge on the dielectric is unknown, and it may vary in time. 
Moreover, considering that the experiment needs to be carried out on time scales of days, the laser has to be switched on and off during the experiment, which may lead to poor stability of the laser beam shape. 
Therefore, a simple observation of the EFISH signal originating in the slowly decaying electric field caused by the surface charge could be a source of systematic errors, as the value of the integral in equation\,\ref{eq_efish} varies in time. 
To avoid these errors, we used the following approach. We applied external voltage on the electrodes, increasing by small steps from zero value to 1-3\,kV, low enough to not cause another breakdown. 
We detected the EFISH signal with the photomultiplier depending on the external voltage. 
All the waveforms were recorded with an oscilloscope, acquired from 64 laser shots. The term ``EFISH signal" ($S$) in the following paragraphs is defined as the depth of the (negative) signal peak from PMT ($\sim P_0^{(2\omega)}$) divided by the square of the peak height of the photodiode monitoring the probe beam power ($\sim P_0^{(\omega)}$),
\begin{equation}\label{eq_S}
    S \propto \frac{P_0^{(2\omega)}}{\left(P_0^{(\omega)}\right)^2}.
\end{equation}
The resulting EFISH signal $S$ revealed a quadratic dependence on the external voltage $V$, with the vertex of the parabola close to the zero value of the signal,
\begin{equation}\label{eq:EFISH_quad}
    S = a \, (V - V_\mathrm{C})^2,
\end{equation}
see Fig.\,\ref{f_efishparaboly}. Let's define the voltage value $V_\mathrm{C}$ with the minimal EFISH signal as the ``compensating voltage". 
For this voltage at the probed coordinate, the electric field originated due to the residual surface charge and the electric field from the externally applied voltage on electrodes combine in a resulting field $E_\mathrm{ext}$ which leads to a zero value of the integral in formula\,\ref{eq_efish}. 
The compensating voltage is therefore a measure of the amount of the residual surface charge and its electric field.
This will be further analyzed and commented in sec.\,\ref{sec:modelresults}.

To study the long-term decay of the surface charge, every set of measurements was performed as follows. 
First, we measured the quadratic signal-to-voltage dependence before the breakdown to ensure that the residual electrical field in the gas gap was close to zero, meaning that the surface charge remaining from the previous experiments was negligible. 
The external voltage was then sharply increased (with a rise time of approximately 1 second, see Fig.\,\ref{fig:scheme_pulse}) beyond the breakdown threshold. This threshold ranged from 4 to 10\,kV, 
depending on the gas type and discharge polarity, resulting in the deposition of surface charges on the dielectrics. 
Just after this ``forward discharge", the second signal-to-voltage dependence was measured in the proximity of the breakdown voltage. This allowed us to determine the compensating voltage corresponding to the charge transferred by the forward breakdown. 
Next, the external voltage was decreased back to zero, causing a reverse/backwards breakdown and transfer of charges with opposite polarity to the dielectric surface causing its effective decrease and thus also decreasing the compensating voltage. 

The backward discharge transferred always smaller amount of electrical charge than the forward discharge. Their difference, the residual surface charge, was then studied, i.e. its decay.

The time of the reverse breakdown (backwards discharge) was marked as $t_0 = 0$. 
After this moment, a series of signal-to-voltage measurements was performed depending on time. 
During every compensating voltage probing, the measurement duration was kept as short as possible (less than one minute) to minimize the effect of the laser beam onto the charge decay. 
The interval between measured points increased from five minutes during the initial times of the experiment to hours at the end. 
The longest measurement series lasted nine days. 
An example of measured signal-to-voltage dependencies is shown in Fig.\,\ref{f_efishparaboly}, showing that the compensating voltage increased from an initial value of 300\,V before the forward discharge to about 3.5\,kV after it, and decreased to 1.6\,kV after the backward discharge.

\section{Numerical model}

Streamer simulations using the fluid approximation were performed \cite{Nijdam_2020, Kulikovsky1997}. 
In particular, a two-fluid (electrons and positive ions) dynamics two-dimensional streamer model similar to the one used in \cite{Tungli2023} was utilized.
Transport of electron and positive ion species was approximated by drift-diffusion equations:
\begin{equation}
    \frac{\partial n_\mathrm{e}}{\partial t} + \nabla \cdot \left( \mu_\mathrm{e} \mathbf{E} \right) - \nabla \cdot D_\mathrm{e} \nabla n_\mathrm{e} = \alpha \mu_\mathrm{e} E n_\mathrm{e} - \beta n_\mathrm{e} n_\mathrm{p}
\end{equation}
\begin{equation}
    \frac{\partial n_\mathrm{pi}}{\partial t} + \nabla \cdot \left( \mu_\mathrm{p} \mathbf{E} \right) - \nabla \cdot D_\mathrm{p} \nabla n_\mathrm{p} = \alpha \mu_\mathrm{e} E n_\mathrm{e} - \beta n_\mathrm{e} n_\mathrm{p}
\end{equation}
where $e,p$ indices stand for electrons and positive ions, respectively, $n$ is the number density, $\mu$ the (signed) mobility, $D$ the diffusivity, $\alpha$ the ionization coefficient, $\beta$ the electron-ion recombination coefficient, $\mathbf{E}$ the electric field and $E$ the magnitude of the electric field.
The transport parameters were approximated using the local field approximation, i.e. they are functions of the electric field magnitude.

The electric field is obtained by taking the negative gradient of electric potential, $\mathbf{E} = -\nabla \phi$, which in turn is obtained by solving Poisson's equation:
\begin{equation}
    \nabla \cdot \left(\varepsilon \nabla \phi \right) = - \sigma \delta_\mathrm{I} - \sum_{i \in \{\mathrm{e},\mathrm{p}\}} q_i n_i
\end{equation}
where $\varepsilon$ is the electric permittivity, $q_i$ is the electric charge of the species $i$ and $\sigma \delta_\mathrm{I}$ is the surface charge density on the solid dielectrics interface.

The simulations performed in this paper are done in an axi-symmetric setup and using the finite volume method.
The Crank-Nicolson scheme with offset coefficient of $0.9$ was used for time-stepping \cite{Crank_Nicolson_1947}.
The advection term was discretized with the van Leer limiter \cite{vanLeer1974}.
The Laplacian terms used linear interpolation for diffusion terms in the transport equation and harmonic interpolation for Poisson's equation.
The gradient of the electric potential is calculated with the least-squares method.
The mesh cell length size target was set to $50$\,$\mu$m and was refined to $0.2$\,mm in front of the solid dielectric to cell length of $1$\,$\mu$m with a growth rate factor of $1.08$.

The cross section for nitrogen were obtained from LxCat \cite{lxcat1,lxcat2,lxcat3}, in particular using the TRINITI data set \cite{lxcat,TRINITI}.
The electron transport properties were calculated using the BOLSIG software \cite{bolsig}.
Electron-ion recombination was adopted from \cite{Kulikovsky1997}, $\beta = 2 \cdot 10^{-13}$\,m$^6$\,s$^{-1}$.
The positive ion mobility value of $3.42$\,cm$^2$/(V $\cdot$ s) was used \cite{Davies1971} and the diffusivity was equal to zero.
The value of $0.1$ was used as the coefficient of electron secondary emission by positive ion impact on the dielectric surface.
The relative permittivity of the solid dielectric was set to $9$.
A pre-ionization density was set in the gas as an initial condition, $n_\mathrm{e}(t=0)~=~n_\mathrm{p}(t=0)~=~10^{9}$\,m$^{-3}$.

\section{Results and discussion}

\subsection{Electrical diagnostics}
\label{sec:electrical}

\begin{figure}[!h]
    \centering
\includegraphics[width=.6\textwidth]{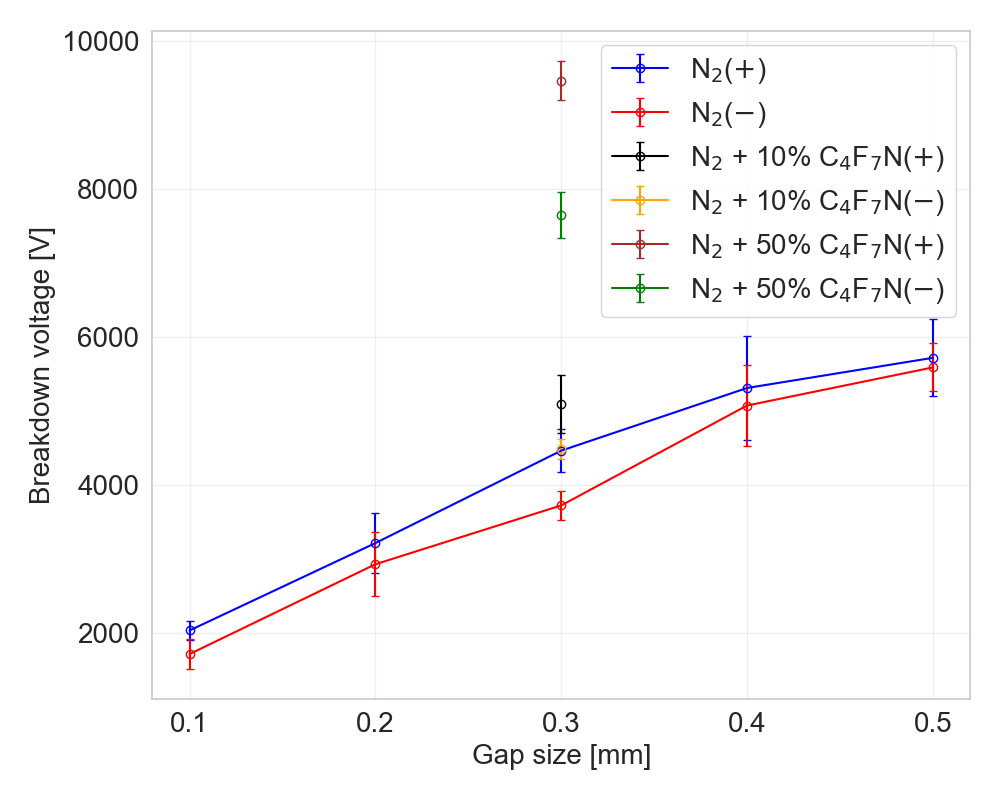}
\caption{Breakdown voltage dependence on different gap sizes for both voltage polarities and three different gas compositions. The absence of data for mixtures with C$_4$F$_7$N for gaps different than 0.3 mm was due to the hardly predictable and non-repetitive behavior of breakdowns.}
\label{breakdown_gap}
\end{figure}

The dependence of breakdown voltage $V_\mathrm{B}$ (externally applied voltage value to the electrode system) on gap size \textit{d} for both voltage polarities is shown in Fig.\,\ref{breakdown_gap}. Note that the breakdown voltages are denoted in absolute values. 
The breakdown voltage is defined as the applied voltage at which the single discharge takes place in the gap. 
From Fig.\,\ref{breakdown_gap} it can be seen  that the breakdown voltage increases with addition of electrically insulating  C$_4$F$_7$N gas. 
This is caused by the strong ability of C$_4$F$_7$N to capture free electrons while at the same time forming negative ions. 
Depletion of free electrons is crucial for the breakdown initiation since the ionization by direct electron impact is the dominant process responsible for the gas breakdown here. 

\begin{figure}[h!]
    \centering
    \includegraphics[width=0.7\linewidth]{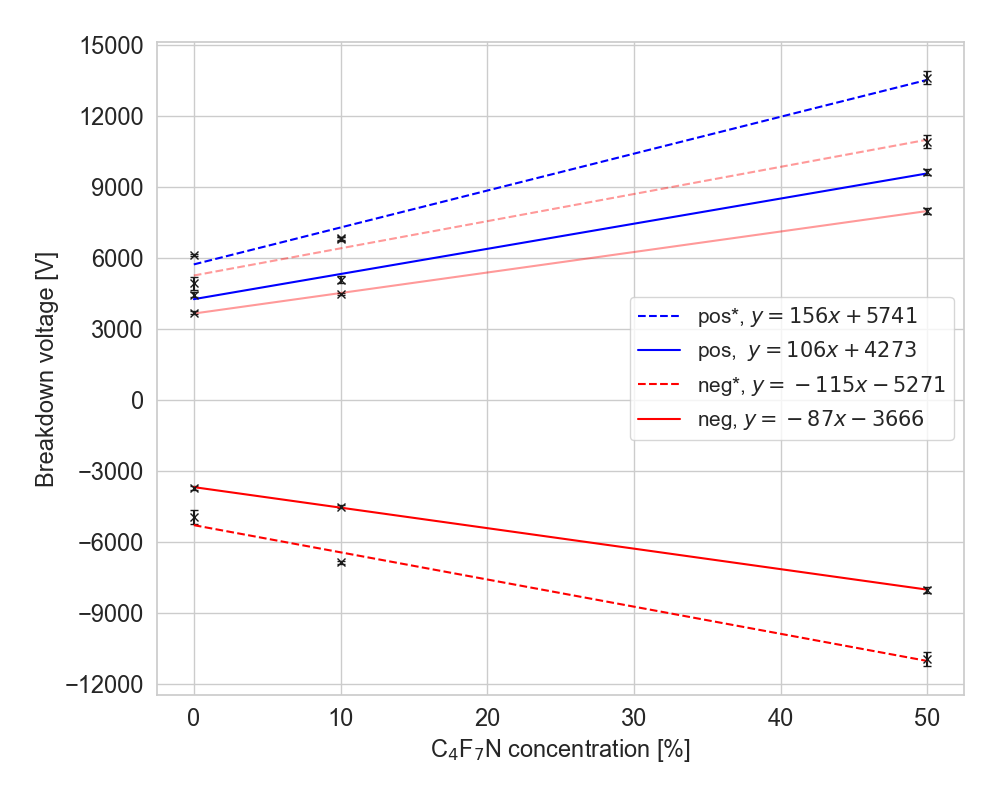} 
\caption{Dependence of breakdown voltage on C$_4$F$_7$N concentration for the case without charge elimination (highlighted by *) and with charge elimination. For better comparison, the negative voltage values were mirrored onto its positive counterpart with lower opacity.}    
\label{boxplot}
\end{figure}

In order to support the EFISH measurements, detailed analysis of current and breakdown voltage measurements was done to provide an independent method of compensating voltage and charge measurements. 
The procedure involved two types of breakdown voltage measurements. 
In the first one, after the deposition of the residual surface charge using a positive or negative polarity (i.e. after forward and backward discharges), a single forward discharge was initiated, and the breakdown voltage was determined. 
In the second one, prior to initiating the discharge, any substantial residual charge amount on the dielectric was eliminated by bringing the electrodes closer together and first then the breakdown voltage was measured. 
The difference in the breakdown voltage for cases with and without surface charge (i.e. without and with the surface charge elimination, respectively) for different C$_4$F$_7$N admixture presented in Fig.\,\ref{boxplot} and table\,\ref{tab:voltage_break}. 
This procedure basically determined the voltage difference, for which is the surface charge responsible in the gap, i.e. an effective analogy of the compensating voltage determined by the EFISH method. 
The $(^\ast)$ in legend highlights the breakdown without prior elimination of the surface charge. The comparison between the results obtained via breakdown voltage analysis and EFISH measurements will be presented in section\,\ref{sec:decay}.

Fig.\,\ref{boxplot} shows that after forward and backward discharge, i.e. after the surface charge deposition, the breakdown voltage  with residual surface charge is higher than with eliminated charge. 
The explanation is that the backward discharge does not erase all the charge deposited during the forward discharge and therefore the remaining charge acts against the next discharge initiation of the same polarity. See also the results presented in Fig.\,\ref{f_efishparaboly}. As it can be seen from Fig.\,\ref{boxplot}, the dependence of the breakdown voltage on the C$_4$F$_7$N admixture is, within some uncertainty,  linear in the investigated range. The equations are given in the figure. 
Apparently, with increasing C$_4$F$_7$N admixture the difference between voltages ($V_\mathrm{B}^\ast - V_\mathrm{B}$) is increasing, clearly showing that statistically, the increased C$_4$F$_7$N admixture causes increase of the residual charge deposited by discharges generated at the breakdown voltage. 

\begin{table}[h!]
\centering
\renewcommand{\arraystretch}{1.4} 
\setlength{\tabcolsep}{14pt}     
\caption{Results of breakdown voltage $V_\mathrm{B}$ with and $V_\mathrm{B}^\ast$ without charge elimination and their difference, i.e. an effective compensating voltage determined via electrical diagnostics. Note that breakdown voltages are denoted in absolute values here.}
\begin{tabular}{lcccc}
\toprule
{Gas (polarity)} & {$V_\mathrm{B}$ [V]} & {$V_\mathrm{B}^\ast$ [V]} & {$V_\mathrm{B}^\ast - V_\mathrm{B}$ [V]} \\ \midrule
N$_{2} (+)$                          & 4500 $\pm$ 300 & 6100  $\pm$ 100 & 1600 $\pm$ 300  \\
N$_{2} (-)$                          & 3700 $\pm$ 200 & 4900  $\pm$ 700 & 1200 $\pm$ 700 \\
N$_{2}$ + 10\% C$_{4}$F$_{7}$N ($+$) & 5100 $\pm$ 400 & 6800  $\pm$ 200 & 1700 $\pm$ 500  \\
N$_{2}$ + 10\% C$_{4}$F$_{7}$N ($-$) & 4500 $\pm$ 100 & 6800  $\pm$ 100 & 2300 $\pm$ 200   \\
N$_{2}$ + 50\% C$_{4}$F$_{7}$N ($+$) & 9500 $\pm$ 300 & 13600 $\pm$ 400 & 4100 $\pm$ 500   \\
N$_{2}$ + 50\% C$_{4}$F$_{7}$N ($-$) & 7700 $\pm$ 300 & 10900 $\pm$ 600 & 3200 $\pm$ 700 \\ \bottomrule
\label{tab:voltage_break}
\end{tabular}
\end{table}

From the breakdown voltage  $V_\mathrm{B}$ (with eliminated surface charge) a value of a breakdown reduced electric field in the gap can be estimated using a one-dimensional approximation \cite{mrkvi2023}:
\begin{equation}
\label{efield}
\frac{E}{N} = \frac{V_B}{N \left(d_{\mathrm{gap}} + \frac{d_{\mathrm{diel}}}{\varepsilon_\mathrm{r}}\right)},
\end{equation}
where $N$ is the neutral gas density, $d_{\mathrm{diel}}$ is the width of dielectric barrier, $d_{\mathrm{gap}}$ is the distance between the electrodes and $\varepsilon_\mathrm{r}$ is the relative permittivity of the dielectrics. The value of a calculated reduced electric field for the case of 0.3 mm gap is shown in table\,\ref{tab:field}.

\begin{table}[h!]
\centering
\renewcommand{\arraystretch}{1.4} 
\setlength{\tabcolsep}{14pt}     
\caption{Reduced electric field strength $E/N$ determined from breakdown voltage in Townsends [Td] and as electric field strength $E$ in [kV/cm].}
\begin{tabular}{lccc}
\toprule
\textbf{Polarity} & \textbf{N$_{2}$} & \textbf{N$_{2}$ + 10\% C$_{4}$F$_{7}$N} & \textbf{N$_{2}$ + 50\% C$_{4}$F$_{7}$N} \\ 
\midrule
\multirow{2}{*}{$(+)$} 
& 440 $\pm$ 30 Td  & 500 $\pm$ 40 Td  & 940 $\pm$ 30 Td \\
& 108 $\pm$ 7 kV/cm  & 124 $\pm$ 9 kV/cm  & 230 $\pm$ 6 kV/cm \\
\midrule
\multirow{2}{*}{$(-)$} 
& 370 $\pm$ 20 Td  & 440 $\pm$ 10 Td  & 760 $\pm$ 30 Td \\
& 90 $\pm$ 5 kV/cm  & 109 $\pm$ 3 kV/cm  & 186 $\pm$ 8 kV/cm \\
\bottomrule
\end{tabular}
\label{tab:field}
\end{table}

The validity of the experimentally determined breakdown $E/N$ values can be assessed by comparing them with theoretical values of a threshold $E/N$ for balance of electron ionization and attachment. 
The ionization and attachment coefficients were computed using the LXCat interface. 
The electron cross sections for nitrogen and C$_{4}$F$_{7}$N mixtures were obtained from LXCat \cite{lxcat1,lxcat2,lxcat3}, using the Biagi data set \cite{Biagi} and Community data set \cite{Community} respectively.
The electron ionization and attachment as a function of $E/N$ were calculated using the BOLSIG software \cite{bolsig}.

\begin{figure}[h!]
    \centering
    \begin{subfigure}[t]{0.49\textwidth}
        \centering
        \includegraphics[width=\textwidth]{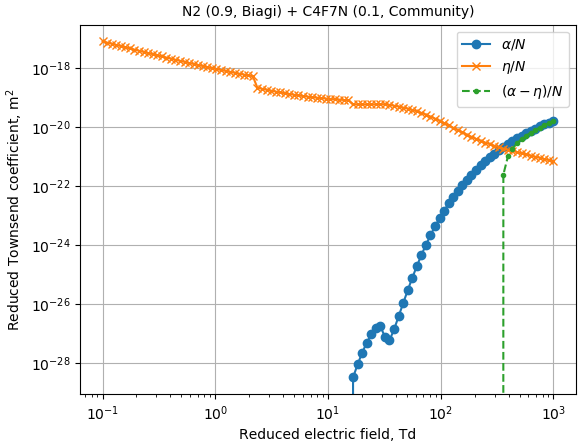} 
        \caption{}
        \label{fig:sub2}
    \end{subfigure}
    \begin{subfigure}[t]{0.49\textwidth}
        \centering
        \includegraphics[width=\textwidth]{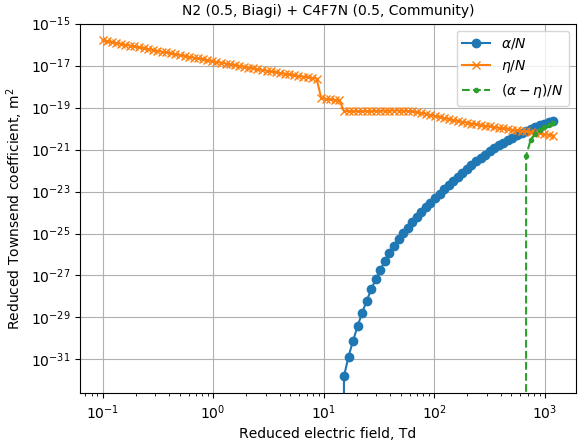} 
        \caption{}
        \label{fig:sub3}
    \end{subfigure}

    \caption{Reduced ionization Townsend coefficient and coefficient of attachment for a), 90 \% N$_2$ + 10 \% C$_{4}$F$_{7}$N, and b), 50 \% N$_2$ + 50 \% C$_{4}$F$_{7}$N.}
    \label{fig:Field}
\end{figure}

From the data shown in Fig.\,\ref{fig:Field} the corresponding critical field values for 10\,\% and 50\,\% admixture of C$_{4}$F$_{7}$N were about 350\,Td and 680\,Td, respectively. Apparently, with increased admixture of highly insulating C$_{4}$F$_{7}$N, the threshold $E/N$ rises. 
These results are coherent with the experimentally determined ones. The difference between the experimental and theoretical threshold electric fields may be accounted to the strong approximation of the equation\,\ref{efield} and to the fact, that probably not all surface charges may be eliminated completely by touching the electrodes. 

\begin{figure}[h!]
    \centering
    \begin{subfigure}{0.3\textwidth}
        \includegraphics[width=\linewidth]{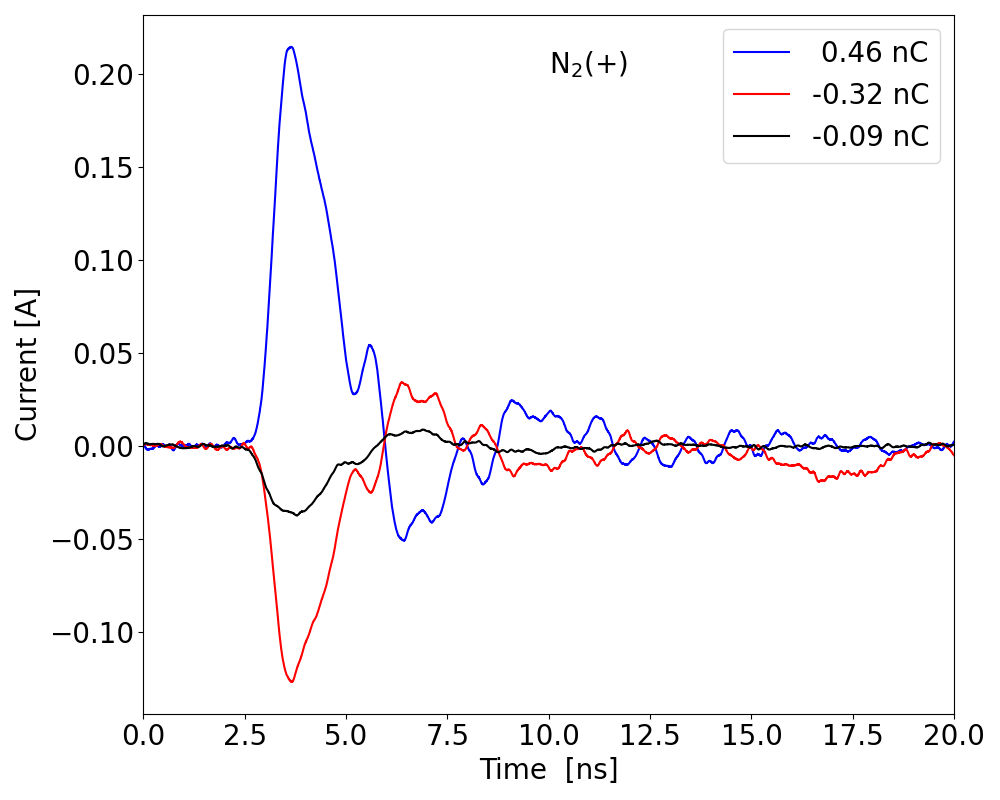} 
        \caption{}
        \label{fig:image1}
    \end{subfigure}
    \hfill
    \begin{subfigure}{0.3\textwidth}
        \includegraphics[width=\linewidth]{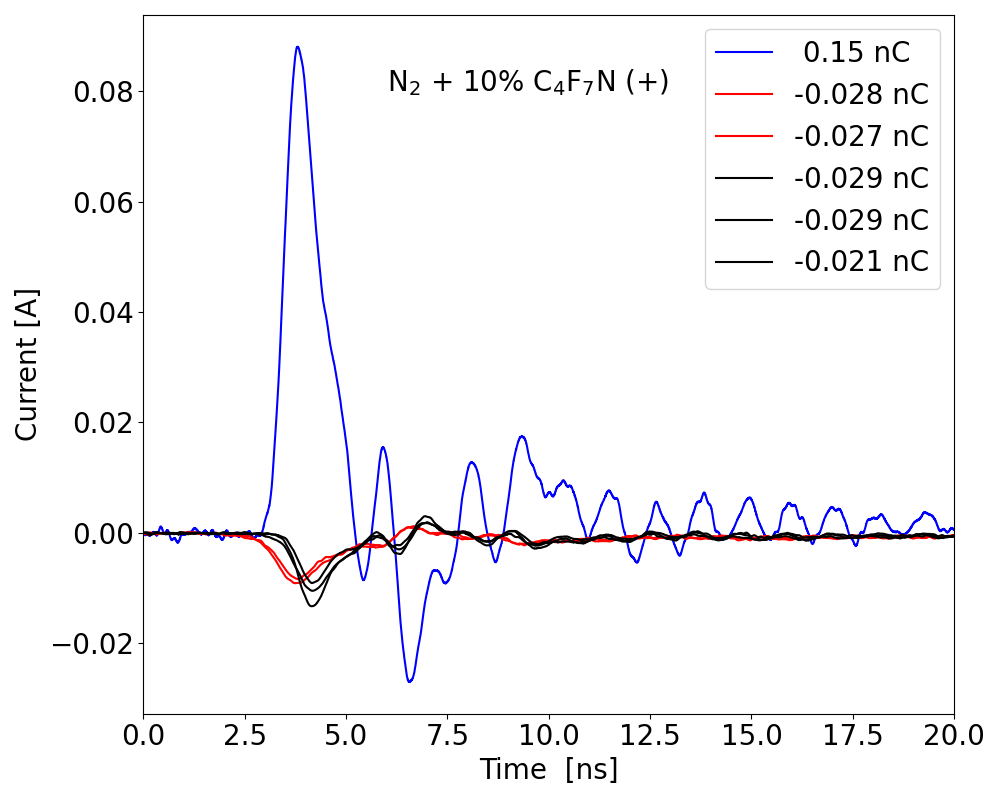} 
        \caption{}
        \label{fig:image2}
    \end{subfigure}
    \hfill
    \begin{subfigure}{0.3\textwidth}
        \includegraphics[width=\linewidth]{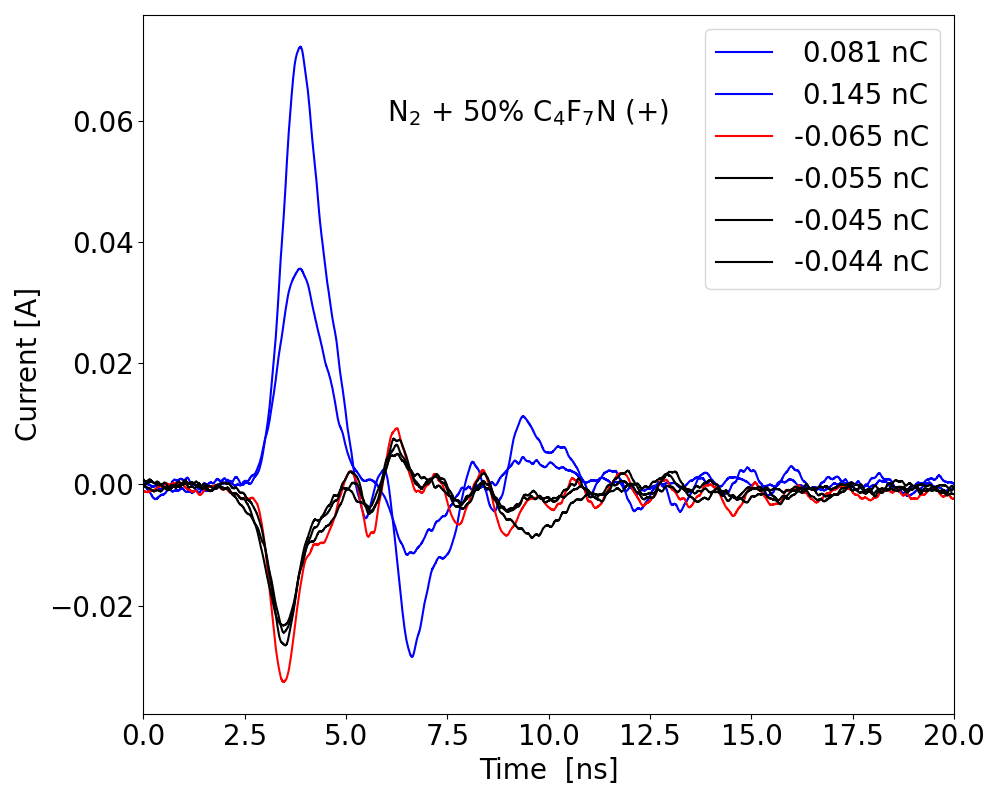} 
        \caption{}
        \label{fig:image3}
    \end{subfigure}
    \begin{subfigure}{0.3\textwidth}
        \includegraphics[width=\linewidth]{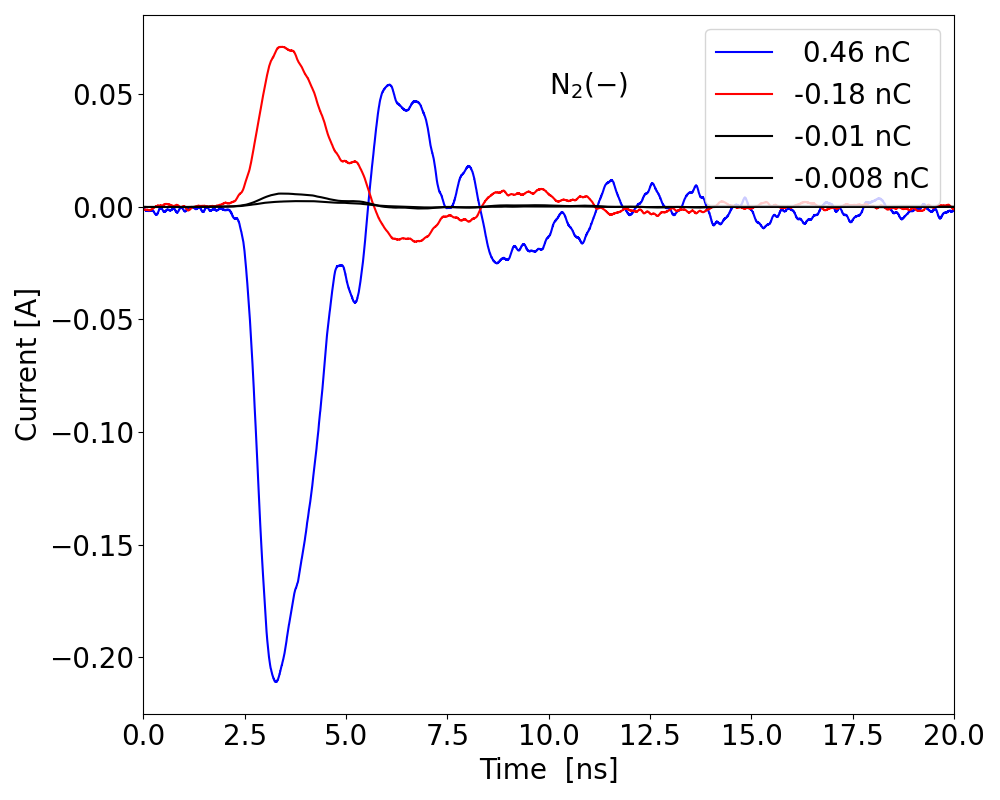} 
        \caption{}
        \label{fig:image1}
    \end{subfigure}
    \hfill
    \begin{subfigure}{0.3\textwidth}
        \includegraphics[width=\linewidth]{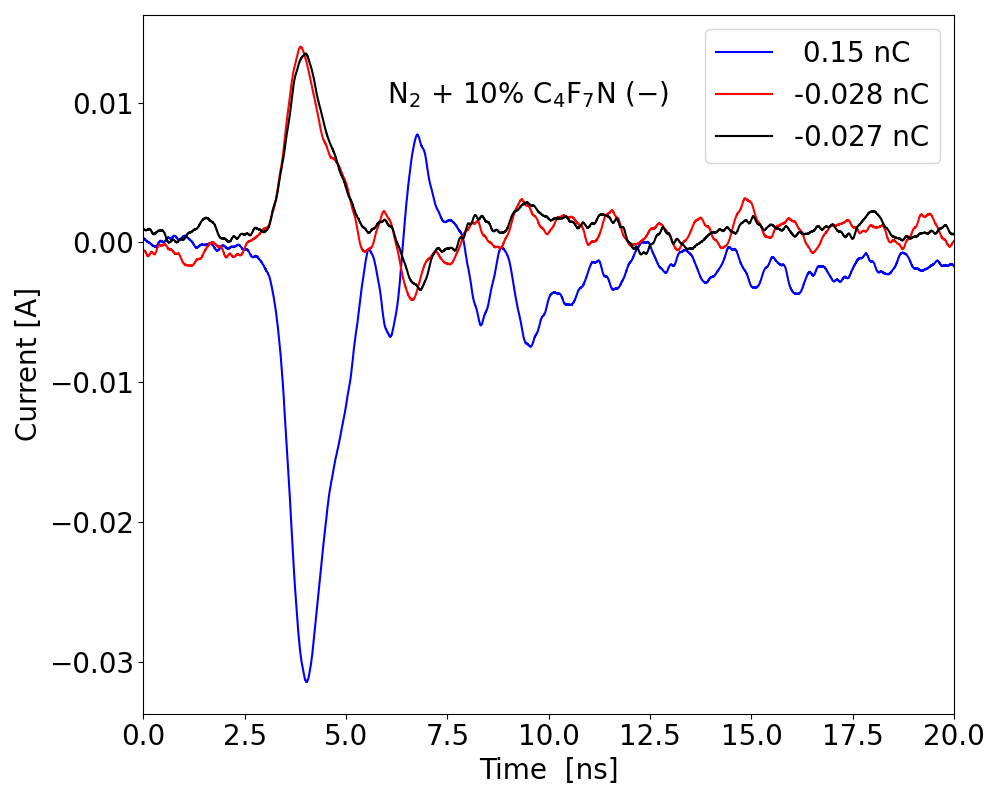} 
        \caption{}
        \label{fig:image2}
    \end{subfigure}
    \hfill
    \begin{subfigure}{0.3\textwidth}
        \includegraphics[width=\linewidth]{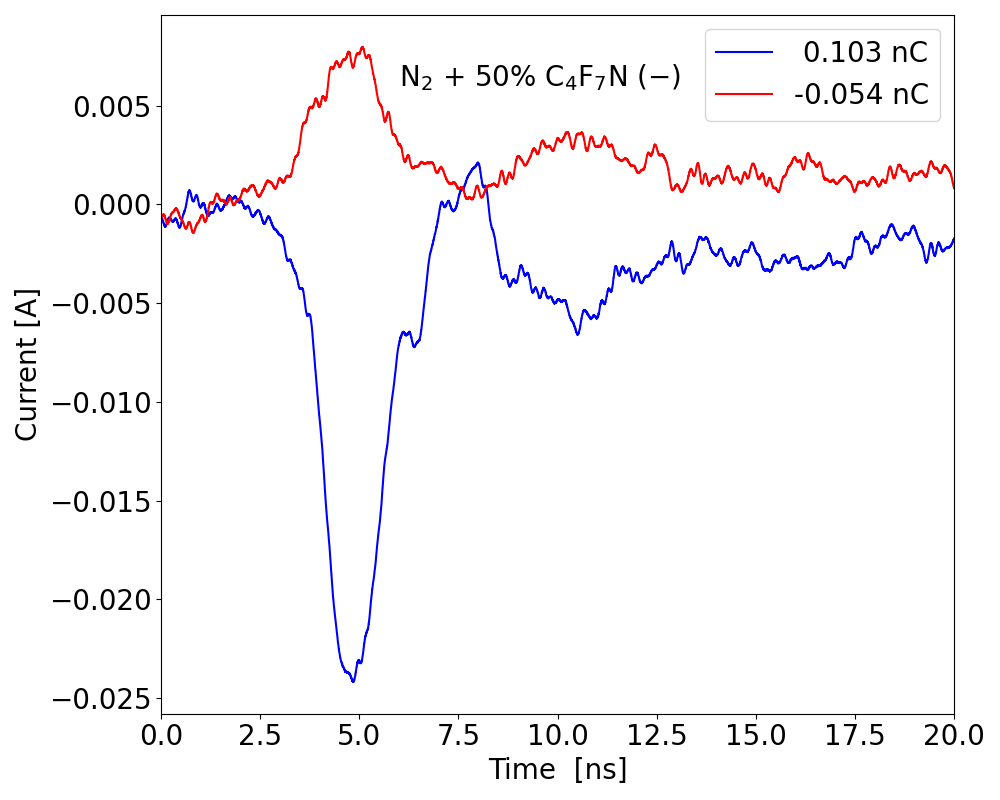} 
        \caption{}
        \label{fig:image3}
    \end{subfigure}     
    \caption{Measured current waveforms for both breakdown polarities,  (a), (b), (c) positive, (d), (e), (f) negative and three gas compositions. Blue - forward discharge, red - backward discharge, black - discharging of residual charges (current pulses recorded as the electrodes were moved closer together to eliminate remaining charge deposited on the dielectric surface). The values of transferred charge have been obtained by integration of the current waveforms taken of its absolute value in the range from 0 to 20\,ns.}
    \label{pulse}
\end{figure}

The electrical current waveforms measured under different conditions are presented in Fig.\,\ref{pulse}. 
At least 10 tests were performed for every gas composition and both voltage polarities to improve the statistical reliability of the measured data. 
Each test/run composed of charge elimination first and then forward and backward discharges, and charge elimination again at the end. 
For breakdowns in pure nitrogen, there was a single forward breakdown (blue curve in the plots) followed by one backward discharge (red curve). 
The discharging of the residual charges (i.e. the elimination of surface charges, given in black curves) for the case of positive polarity happened instantly as the electrodes moved closed together. 
In the case of the negative polarity, the discharging of the residual charge on dielectrics happened in more than one step. 
This effect of multiple breakdowns was even more visible in mixtures with higher content of C$_4$F$_7$N. 

From the measured current waveforms, the mean values of full width at half maximum (FWHM), rise time of the current pulses and an average value of transferred charge for both forward (Q$_\mathrm{F}$) (see table\,\ref{forwardpulse}) and backward discharge (Q$_\mathrm{B}$) (see table\,\ref{backwardpulse}) were determined. 
Note that the rise times of the discharges are under studied conditions in far sub-nanosecond range. 
For some cases, even the FWHM is in such ultra-short range. 

The value of the deposited residual charge, whose decay is further studied by the EFISH technique, is given by (Q$_\mathrm{F}$ - Q$_\mathrm{B}$) (see table\,\ref{chargebalance}). 
This charge amount is decreasing with the increasing admixture of C$_4$F$_7$N - this trend will be discussed also later with respect to the compensating voltage measurements. 
For the positive and negative polarity discharge in pure nitrogen the value was Q$_\mathrm{F}$ - Q$_\mathrm{B}$ = (140$ \pm$40)\,pC and Q$_\mathrm{F}$ - Q$_\mathrm{B}$ = (390$ \pm$50)\,pC respectively. 
This values will be used later for comparison with results of numerical modeling in the same gas. 

In the last column of the table\,\ref{chargebalance}, the charge transferred during residual charge elimination process (by bringing the electrodes closer) is also subtracted to see the total balance. 
The total balance should be zero in ideal case. However, the value of Q$_\mathrm{F}$-Q$_\mathrm{B}$-$\Sigma$Q$_\mathrm{Res}$ is far from zero and also relatively high compared to the values of Q$_\mathrm{F}$-Q$_\mathrm{B}$. 
Therefore, the conclusion is that the value of Q$_\mathrm{F}$-Q$_\mathrm{B}$-$\Sigma$Q$_\mathrm{Res}$ represents relatively high uncertainty of these current measurements. The uncertainty has two reasons, both related to the secondary discharges. 
First, even if the electrodes were in direct contact, it does not have to lead to discharging of all the residual charge domains on the dielectric surface. The fact that the surface charge is deposited inhomogeneously and spreads over the surface is known \cite{zhu1996}. 
Second, possibly not all the backward and charge-eliminating discharges may be detected, as their amplitudes decreased to units of milliamperes while the pulse widths were in sub-nanosecond scale which is generally a challenging task to measure.

\begin{table}[h!]
\centering
\renewcommand{\arraystretch}{1.4} 
\setlength{\tabcolsep}{14pt}     
\caption{Selected characteristics calculated from the measured electrical current for forward discharge.}
\begin{tabular}{lcccc}
\toprule
{Gas (voltage polarity)} & {FWHM [ns]} & {Rise Time [ns]} & {Q$_\mathrm{F}$[nC]} \\ \midrule
N$_{2} (+)$ & 1.42 $\pm$ 0.07 & 0.55 $\pm$ 0.02 & 0.43 $\pm$ 0.10  \\
N$_{2} (-)$ & 1.31 $\pm$ 0.08 & 0.53 $\pm$ 0.01 & 0.62 $\pm$ 0.10  \\
N$_{2}$ + 10\% C$_{4}$F$_{7}$N ($+$) & 0.92 $\pm$ 0.07 & 0.50 $\pm$ 0.04 & 0.18 $\pm$ 0.03 \\
N$_{2}$ + 10\% C$_{4}$F$_{7}$N ($-$) & 1.15 $\pm$ 0.19 & 0.65 $\pm$ 0.11 & 0.19 $\pm$ 0.01 \\
N$_{2}$ + 50\% C$_{4}$F$_{7}$N ($+$) & 1.20 $\pm$ 0.20 & 0.71 $\pm$ 0.15 & 0.13 $\pm$ 0.02  \\
N$_{2}$ + 50\% C$_{4}$F$_{7}$N ($-$) & 1.30 $\pm$ 0.30 & 0.90 $\pm$ 0.30 & 0.12 $\pm$ 0.02 \\ \bottomrule
\label{forwardpulse}
\end{tabular}
\end{table}

\begin{table}[h!]
\centering
\renewcommand{\arraystretch}{1.4} 
\setlength{\tabcolsep}{14pt}     
\caption{Selected characteristics calculated from the measured electrical current for backward discharge.}
\begin{tabular}{lcccc}
\toprule
{Gas (voltage polarity)} & {FWHM [ns]} & {Rise Time [ns]} & {Q$_\mathrm{B}$[nC]} \\ \midrule
N$_{2} (+)$ & 1.57 $\pm$ 0.08 & 0.61 $\pm$ 0.03 & 0.29 $\pm$ 0.09 \\
N$_{2} (-)$ & 1.62 $\pm$ 0.03 & 0.65 $\pm$ 0.01 & 0.22 $\pm$ 0.04 \\
N$_{2}$ + 10\% C$_{4}$F$_{7}$N ($+$) & 1.66 $\pm$ 0.09 & 0.91 $\pm$ 0.06 & 0.05 $\pm$ 0.01 \\
N$_{2}$ + 10\% C$_{4}$F$_{7}$N ($-$) & 1.27 $\pm$ 0.06 & 0.57 $\pm$ 0.04  & 0.04 $\pm$ 0.01 \\
N$_{2}$ + 50\% C$_{4}$F$_{7}$N ($+$) & 1.41 $\pm$ 0.11 & 0.71 $\pm$ 0.08 & 0.06 $\pm$ 0.01 \\
N$_{2}$ + 50\% C$_{4}$F$_{7}$N ($-$) & 1.83 $\pm$ 0.54 & 0.69 $\pm$ 0.09 & 0.07 $\pm$ 0.03 \\ \bottomrule
\label{backwardpulse}
\end{tabular}
\end{table}

\begin{table}[h!]
\centering
\renewcommand{\arraystretch}{1.4} 
\setlength{\tabcolsep}{14pt}     
\caption{Calculated charge balance after forward/backward discharging and forward/backward/charge elimination process.}
\begin{tabular}{lcccc}
\toprule
{Gas (voltage polarity)} & {Q$_\mathrm{F}$-Q$_\mathrm{B}$\,[nC]} & {Q$_\mathrm{F}$-Q$_\mathrm{B}$-$\Sigma$Q$_\mathrm{Res}$\,[nC] }  \\ \midrule
N$_{2} (+)$ & 0.14 $\pm$ 0.04 & 0.03 $\pm$ 0.03  \\
N$_{2} (-)$ & 0.40 $\pm$ 0.05 & 0.36 $\pm$ 0.06  \\
N$_{2}$ + 10\% C$_{4}$F$_{7}$N ($+$) & 0.13 $\pm$ 0.03 & 0.016 $\pm$ 0.01  \\
N$_{2}$ + 10\% C$_{4}$F$_{7}$N ($-$) & 0.15 $\pm$ 0.01 & 0.095 $\pm$ 0.05   \\
N$_{2}$ + 50\% C$_{4}$F$_{7}$N ($+$) & 0.07 $\pm$ 0.02 & 0.017 $\pm$ 0.007  \\
N$_{2}$ + 50\% C$_{4}$F$_{7}$N ($-$) & 0.05 $\pm$ 0.03 & 0.049 $\pm$ 0.03  \\ \bottomrule
\label{chargebalance}
\end{tabular}
\end{table}

\subsection{Determination of the C$_4$F$_7$N hyperpolarizability}

The EFISH measurements performed in three different gas mixtures allow us to estimate the value of the third-order nonlinear hyperpolarizability $\alpha^{(3)}$ of C$_4$F$_7$N gas. In particular, we will determine the component $\alpha^{(3)}_\parallel(-2\omega, 0, \omega, \omega)$ for the wavelength 1064 nm ($\omega = 1.77\cdot10^{15}$ s$^{-1}$). 
Assuming that both the spatial distribution of the electric field $E_\mathrm{ext}$ and the laser beam shape undergo negligible changes during the measurements, the EFISH signal $S$ from formulas\,\ref{eq_efish} and \ref{eq_S} can be simplified to
\begin{equation}
    S = \left[ \alpha^{(3)} \cdot N \right]^2 \cdot E_0^2 \cdot k_0, 
\end{equation}
where $k_0$ is constant, characterizing the interaction of the laser with the electric field. In the case of using a mixture of gases, the nitrogen and C$_4$F$_7$N of concentrations $N_\mathrm{N_2}$ and $N_\mathrm{ C_4F_7N}$, respectively, the signal will be affected by both different hyperpolarizabilities of these gases:
\begin{equation}
    S = \left[ \alpha^{(3)}_{\mathrm{N_2}} \cdot N_{\mathrm{N_2}} +  \alpha^{(3)}_{\mathrm{ C_4F_7N}} \cdot N_{\mathrm{C_4F_7N}} \right]^2 \cdot E_0^2 \cdot k_0. 
\end{equation}
If the total concentration $N_\mathrm{total} = N_\mathrm{C_4F_7N} + N_\mathrm{N_2}$ is constant, the equation can be rewritten to
\begin{equation}
    S = k_0 \cdot \left(\alpha^{(3)}_{\mathrm{N_2}} \cdot N_\mathrm{total} \right)^2 \cdot
    \left[ \left(  \frac{\alpha^{(3)}_{\mathrm{ C_4F_7N}}}{\alpha^{(3)}_{\mathrm{N_2}}}-1 \right) \frac{N_{\mathrm{C_4F_7N}}}{N_{\mathrm{total}}} + 1
    \right]^2 \cdot E_0^2. 
\end{equation}
In our experiment, we varied the magnitude of the electric field $E_0$ via the addition of an external voltage $V$ on electrodes. Then the coefficient $a$ from parabolic dependence $S = a (V-V_\mathrm{C})^2$ is defined as
\begin{equation}\label{eq:Aalfa}
    a = k \cdot \left( \alpha^{(3)}_{\mathrm{N_2}} \cdot N_\mathrm{total} \right)^2 \cdot
    \left[ \left(  \frac{\alpha^{(3)}_{\mathrm{ C_4F_7N}}}{\alpha^{(3)}_{\mathrm{N_2}}}-1 \right) \frac{N_{\mathrm{C_4F_7N}}}{N_{\mathrm{total}}} + 1
    \right]^2.
\end{equation}
When plotting the coefficient $a$ for different C$_4$F$_7$N contents in the gas $\left( \frac{N_{\mathrm{C_4F_7N}}}{N_{\mathrm{total}}} \right)$, the dependence is also parabolic, with only two independent parameters: $k \cdot \left( \alpha^{(3)}_{\mathrm{N_2}} \cdot N_\mathrm{total} \right)^2$ and $\frac{\alpha^{(3)}_{\mathrm{C_4F_7N}}}{\alpha^{(3)}_{\mathrm{N_2}}}$. 

\begin{figure}[htbp]
    \centering
    \begin{subfigure}[t]{0.49\textwidth}
        \centering
        \includegraphics[width=\textwidth]{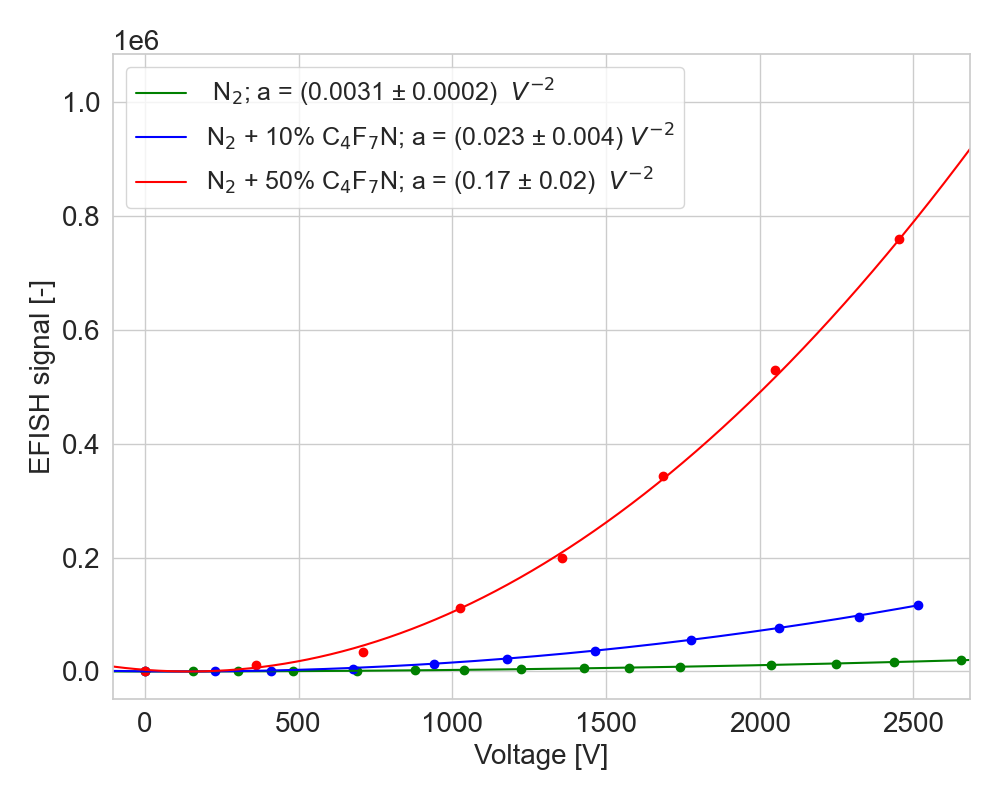} 
        \caption{}
        \label{fig:sub2}
    \end{subfigure}
    \begin{subfigure}[t]{0.49\textwidth}
        \centering
        \includegraphics[width=0.8\textwidth]{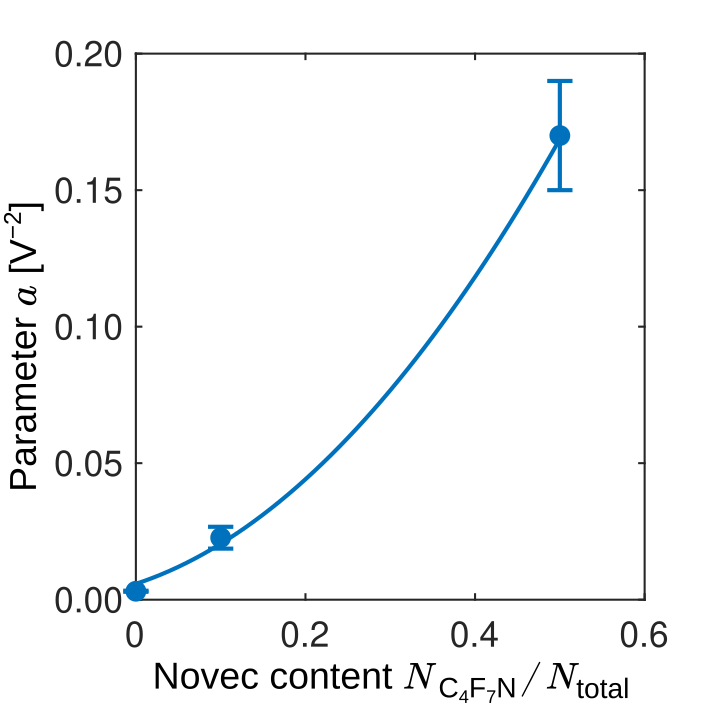} 
        \caption{}
        \label{fig:sub3}
    \end{subfigure}
\caption{a) Example of EFISH signal-voltage dependencies obtained in three different gas mixtures (pure nitrogen, 10 \% C$_4$F$_7$N + 90 \% N$_2$, 50 \%  C$_4$F$_7$N + 50 \% N$_2$), b) parameters $a$ from the parabolic dependence $S = a(V-V_\mathrm{C})^2$ fitted by equation\,\ref{eq:Aalfa}.}    
\label{fig:hyperalfa}
\end{figure}

Fig.\,\ref{fig:hyperalfa} a) shows an example of signal-voltage dependencies for three different C$_4$F$_7$N contents in the gas. 
To obtain parameter $a$ for each content, we extracted the parameters from all measurements with a given concentration and calculated their mean value and standard deviation: $a_\mathrm{N_2} = (0.0031\pm0.0002)$ V$^{-2}$, $a_\mathrm{10\%\,C_4F_7N} = (0.023\pm0.004)$ V$^{-2}$, $a_\mathrm{50\%\,C_4F_7N} = (0.17\pm0.02)$ V$^{-2}$.  
As all the presented charge decay measurements were accounted, the standard deviations of $a$ also  include the impact of possible changes in electric field distribution $E_{ext}(z)$ or the variations in the laser beam shape thorough the entire experiment on the measured EFISH signal.
In Fig.\,\ref{fig:hyperalfa} b), these values are fitted with formula\,\ref{eq:Aalfa}, getting the ratio of C$_4$F$_7$N and nitrogen hyperpolarizabilities
$\frac{\alpha^{(3)}_{\mathrm{C_4F_7N}}}{\alpha^{(3)}_{\mathrm{N_2}}} = (12 \pm 3)$. 
Taking the literature value for nitrogen, $\alpha^{(3)}_\mathrm{N_2, \parallel} (-2\omega; 0, \omega, \omega)= (60.1 \pm 0.2)\cdot 10^{-63}$ C$^4$$\cdot$m$^{4}\cdot$J$^{-3}$ \cite{Shelton1990}, resulting hyperpolarizability of C$_4$F$_7$N would reach value $\alpha^{(3)}_\mathrm{C_4F_7N, \parallel} (-2\omega; 0, \omega, \omega)= (7.2 \pm 1.8)\cdot 10^{-61}$ C$^4$$\cdot$m$^{4}\cdot$J$^{-3}$.

\subsection{Field homogeneity test}

In this section, the dependence of the EFISH signal on the position $x$ in the inter-electrode gap is investigated. 
For different positions $x$, the measured signal $S(x,V)$ still shows a parabolic dependence on the externally applied voltage $V$:
\begin{equation}
    S(x,V) = a(x) \cdot (V-V_\mathrm{C})^2.
\end{equation}
The parameter $a(x)$ is in theory spatially dependent as it contains information about the spatial shape of the electric field $E_\mathrm{ext}(x, z)$, i.e. how the electric field (combined both from surface charge and external voltage) distribution along the $z$ axis varies along the position in the gap $x$.

The reactor was placed on the linear translation stage allowing the precise movement of the full reactor chamber with respect to the laser beam. The EFISH signal-voltage dependencies were recorded for different positions in the gap. The measurements started at position $0.10$\,mm, where all the charge decay experiments were performed. 
We then moved closer to the metallic electrode in steps of 20\,$\mu$m up to the position $0.26$\,mm. After this point, we returned to positions closer to the dielectric electrode, i.e. positions 0.12--0.02\,mm.

\begin{figure}[!h]
    \centering
\includegraphics[width=.7\textwidth]{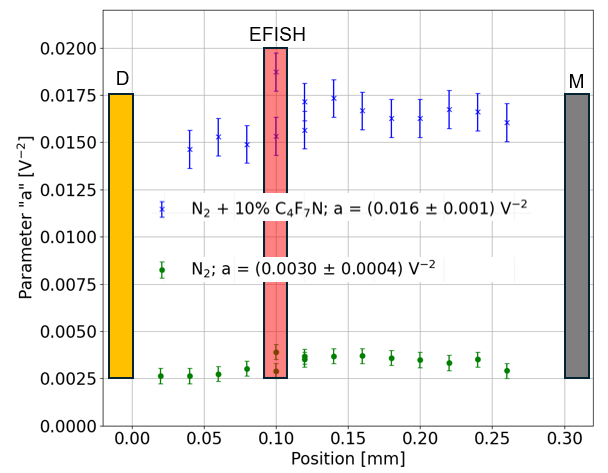}
\caption{Values of parameter ``$a$" measured across the discharge gap. Position for charge decay measurements is highlighted by red rectangle. The position of the dielectric and metal electrode were highlighted by orange and grey rectangle respectively.}
\label{f:Homogen}
\end{figure}

In Fig.\,\ref{f:Homogen}, the resulting parameters $a(x)$ are plotted for all the positions $x$ and two gas mixtures: pure N$_2$ and 10\% C$_4$F$_7$N + 90\% N$_2$. The change in the laser beam shape led to a gradual decrease in the value of $a$ of about 10\% in time; otherwise, the value of $a$ was constant within all measured positions. This experiment confirms the homogeneity of the $x$-axis component of the electric field within gap close the the axis.
A practical consequence of this finding is that the presented measurements are not sensitive to the precise positioning of the $x$ coordinate of the reactor (which had to be moved outside the laboratory and back when it was refilled by a new C$_4$F$_7$N mixture).

\subsection{Surface charge decay}
\label{sec:decay}
\begin{figure}[h!]
    \centering
    \begin{subfigure}{0.49\textwidth}
        \includegraphics[width=\linewidth]{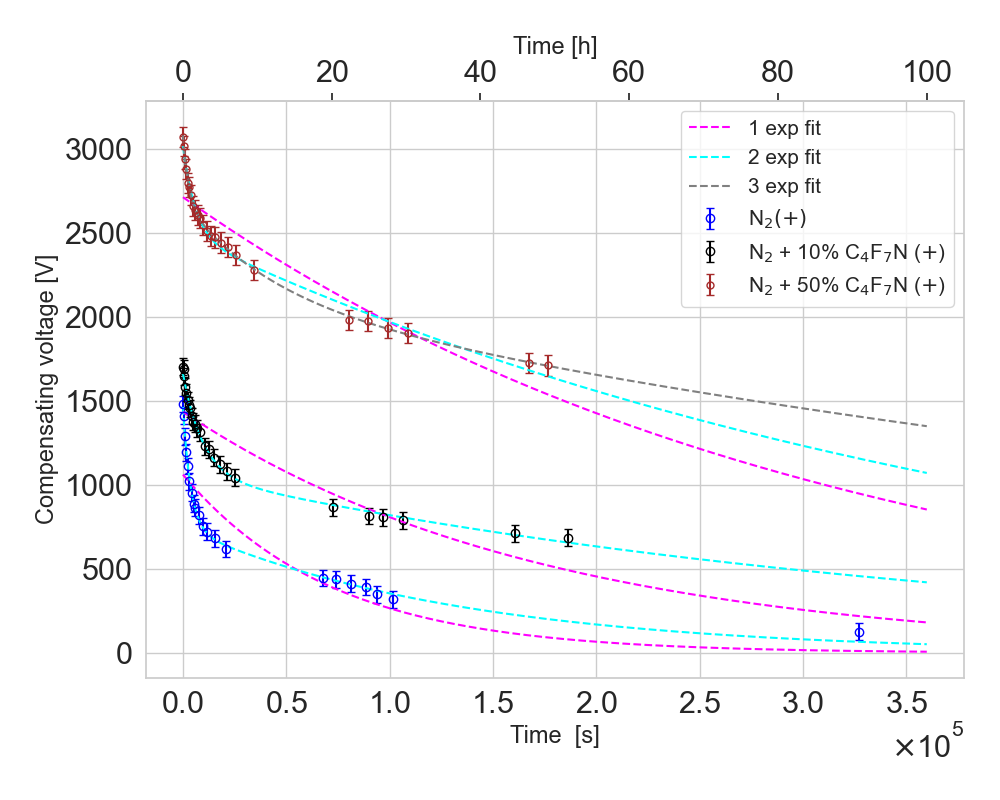} 
        \caption{}
        \label{fig:image1}
    \end{subfigure}
    \begin{subfigure}{0.49\textwidth}
        \includegraphics[width=\linewidth]{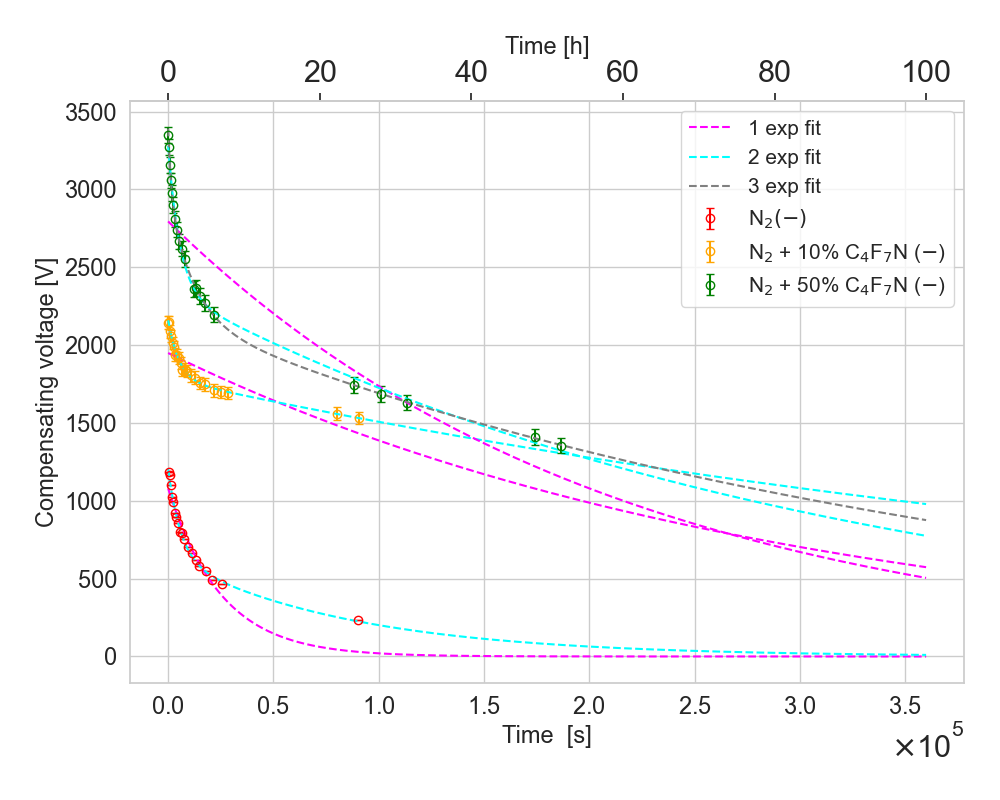} 
        \caption{}
        \label{fig:image1}
    \end{subfigure}
\caption{Compensating voltage decay in pure N$_2$ and in N$_2$ with 10  \% and 50 \% C$_4$F$_7$N admixture. For a) positive polarity, b) negative polarity.}
\label{exp_fit}
\end{figure}

The surface charge decay over time was monitored through the compensating voltage, the external voltage that balances with the electric field induced by the deposited charge, resulting in the minimum EFISH signal, 
as described in section\,\ref{methodology}. 
The decay of the surface charge is reflected in the decay of the compensating voltage values. The results of the EFISH measurements are shown in Fig.\,\ref{exp_fit}. 
Note that all the compensating voltages are shown as absolute values, i. e. for the negative polarity discharges all the data are reversed. 
Apparently, the compensating voltage is decreasing on the timescale of hours and even days after its deposition. 
The polarity sign describes the high-voltage at the metalic electrode. 
For a positive pulse (polarity $+$) during forward discharge, therefore, the deposited charges on the opposite, i.e., grounded and dielectric-covered, electrode should have positive polarity, i.e. positive ions. For the negative polarity, the negative ions and electrons are deposited onto the dielectrics.

\begin{figure}[!h]
    \centering
    a)
\includegraphics[width=.45\textwidth]{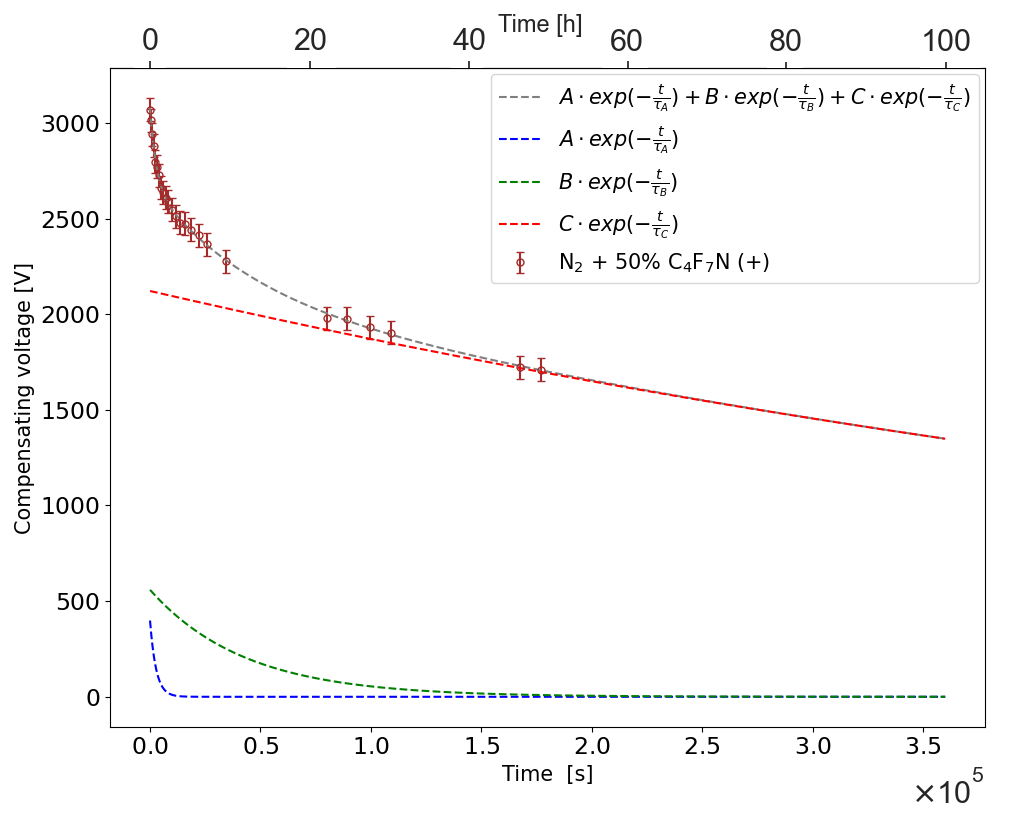}
b)
\includegraphics[width=.46\textwidth]{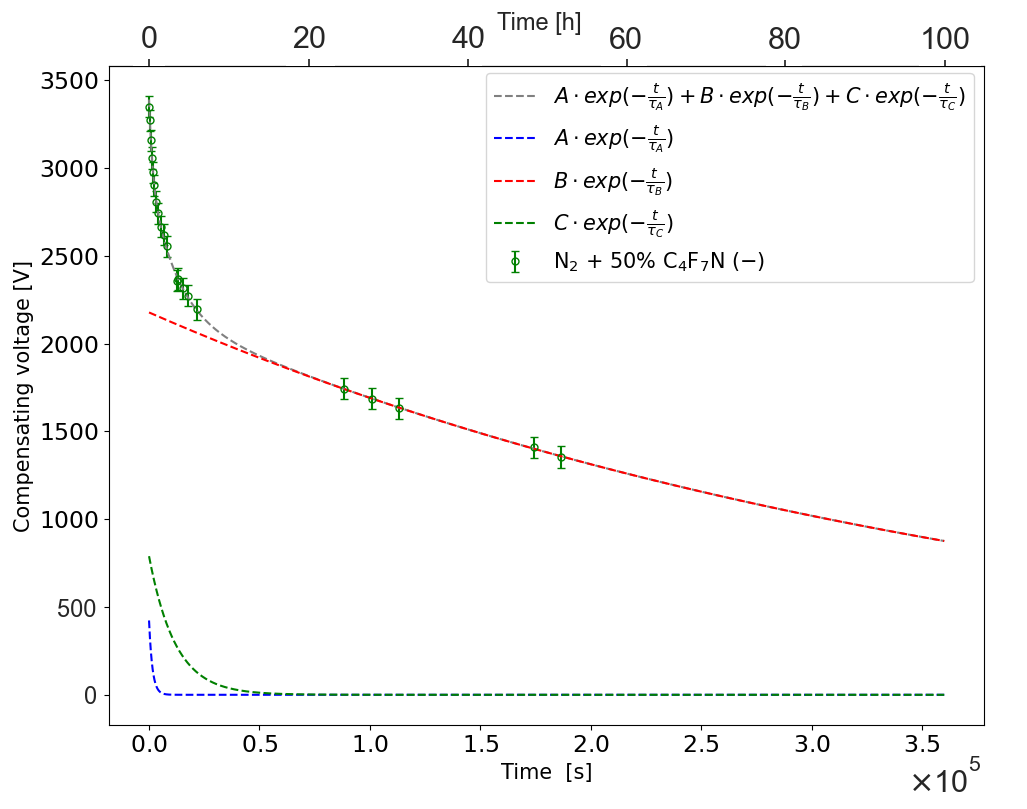}
\caption{Individual components of eq.\,(\ref{exp}) for the N$_2$ with 50 \% C$_4$F$_7$N admixture. For a) positive polarity, b) negative polarity.}
\label{decay3}
\end{figure}

Charge decay is overall a complex process and simple exponential decay modeling may not always be sufficient  \cite{molinie2024}. 
Nevertheless, for the first analyses of the experimental results we have decided to represent the measured data by a sum of two or three simple exponential decay functions, obtaining a good agreement in the fit. 
In Fig.\,\ref{exp_fit}, the compensating voltage data is fitted using single, double, or triple exponential curves, represented by the following formula:

\begin{equation}
\label{exp}
V_\mathrm{C} (t) = A \mathrm{e}^{-\frac{t}{\tau_\mathrm{A}}} + B \mathrm{e}^{-\frac{t}{\tau_\mathrm{B}}} + C \mathrm{e}^{-\frac{t}{\tau_\mathrm{C}}}.
\end{equation}

In general, the best fit was achieved using double and triple exponential models. This result suggests that the decay of the charge occurs on multiple time scales, having two or three physical reasons. 
In particular, for the case with pure nitrogen, there was little or no difference with the use of a double or triple exponential model. 
A similar situation happened for nitrogen with 10 \% C$_4$F$_7$N admixture. 
However, for the situation with 50 \% C$_4$F$_7$N admixture the decay of the charge appears to be more complicated since it requires three exponential fit to better characterize the charge decay data. The contribution of the three exponential terms to the charge decay is shown in Fig.\,\ref{decay3}.

\begin{figure}[!h]
    \centering
    a)
\includegraphics[width=.47\textwidth]{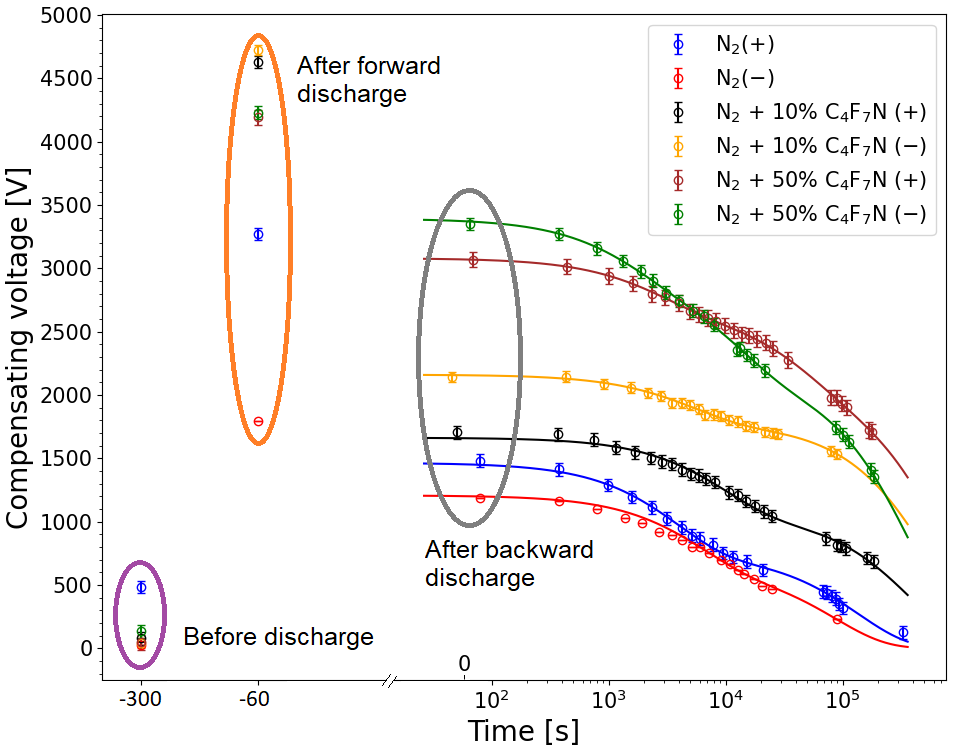}
b)
\includegraphics[width=.45\textwidth]{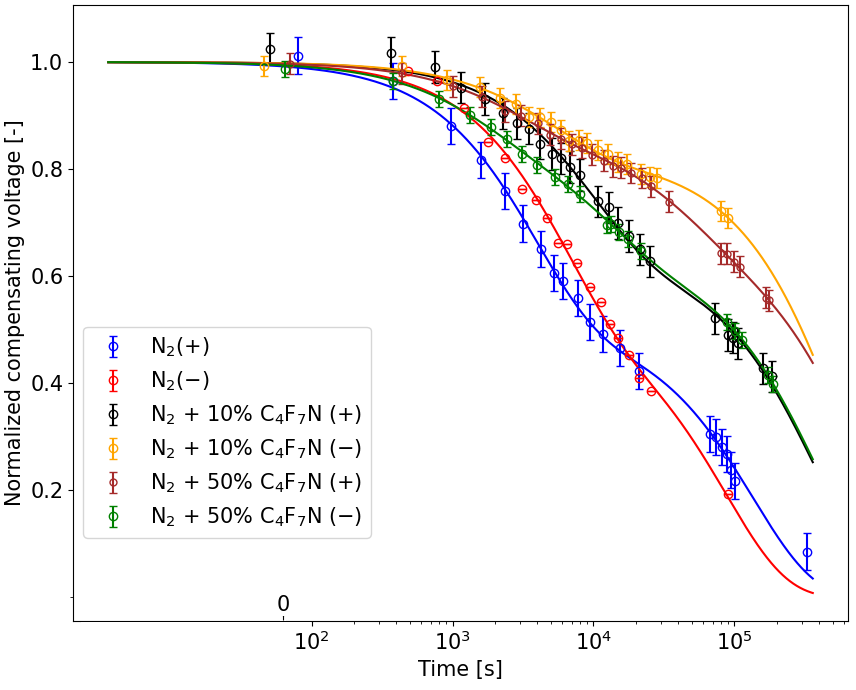}
\caption{Charge decay on a dielectric surface in absolute values, see part a). Purple ellipse - state before the discharge ($t \approx -5$\,min). Orange ellipse - after forward discharge ($t \approx -1$\,min). Grey ellipse - after backward discharge when the actual charge decay measurement starts. All curves are fitted by sum of two exponential functions except for the admixture of 50\% C$_{4}$F$_{7}$N which uses a sum of three exponential functions, see equation\,\ref{exp}. In part b), the normalized curves are presented.}
\label{decay}
\end{figure}

The measured data for charge decay after voltage pulses of both polarities and in all admixtures are shown together in Fig.\,\ref{decay}. 
The time scales are shown in seconds and hours. 
In part a), the compensating voltage in absolute values is shown, while in part b), the normalized values are presented. 
In part a), prior to the charge decay, the process of the dielectric surface charging is visible on the compensating voltage values. First, the compensating voltage for the surface after charge elimination, i.e. before forward and backward discharges, is measured. 
As one can see in the values denoted as ``Before discharge", the elimination of the charge by touching the electrode surfaces is not always perfect. 
This fact can influence the values and distribution of the deposited charge in the next steps, i.e. forward and backward discharges. 
However, the charge decay is then measured with respect to the value after these two discharges, denoted as ``After backward discharge".

\begin{table}[h!]
\centering
\renewcommand{\arraystretch}{1.4} 
\setlength{\tabcolsep}{14pt}     
\caption{Compensating voltage determination via electrical diagnostics, $V_\mathrm{B}^\ast - V_\mathrm{B}$ (results from 10 measurements for each mixture and polarity) and EFISH, $V_\mathrm{C}$ (results of single measurement), both after the backward discharge.} 
\begin{tabular}{lccc}
\toprule
{Gas (polarity)}  & {$V_\mathrm{B}^\ast - V_\mathrm{B}$}\,[V] & {$V_\mathrm{C}$}\,[V] \\ \midrule
N$_{2} (+)$                           & 1600 $\pm$ 300 & 1500 $\pm$ 50  \\
N$_{2} (-)$                          & 1200 $\pm$ 700 & 1190 $\pm$ 50 \\
N$_{2}$ + 10\% C$_{4}$F$_{7}$N ($+$)  & 1700 $\pm$ 500 & 1700 $\pm$ 50  \\
N$_{2}$ + 10\% C$_{4}$F$_{7}$N ($-$)  & 2300 $\pm$ 200 & 2140 $\pm$ 40  \\
N$_{2}$ + 50\% C$_{4}$F$_{7}$N ($+$)  & 4100 $\pm$ 500 & 3070 $\pm$ 60  \\
N$_{2}$ + 50\% C$_{4}$F$_{7}$N ($-$)   & 3200 $\pm$ 700 & 3350 $\pm$ 50  \\ \bottomrule
\label{tab:voltage_comp}
\end{tabular}
\end{table}

Table \ref{tab:voltage_comp} shows the compensating voltages $V_C$ just after the charging process (charge elimination and following forward and backward discharges) determined from single runs of EFISH measurements for both polarities and all investigated mixtures. The data are compared with corresponding effective compensating voltage values V$_B$*-V$_B$ determined statistically from the electrical breakdown voltage measurements as discussed in \ref{sec:electrical}. 
While the measurements generally align within the error margins of our experimental capabilities, the compensating voltage determined via EFISH exhibits a tendency to be lower than the corresponding values obtained through electrical diagnostics. 
This discrepancy will be analyzed and addressed in greater detail in the following section with help of numerical modeling. 

From the data in table \ref{tab:voltage_comp} one can see that with one deviation, it can be concluded that both methods arrive coherently to the same conclusion, i.e. with increasing C$_4$F$_7$N admixture the amount of the residual charge deposited within elimination/forward/backward discharge process grows. 
This finding is not in agreement with the results presented earlier in table \ref{chargebalance}, where residual charge was determined from measured electrical currents. 
Apparently, the electrical current measurements are loaded with high uncertainty, as discussed earlier, and may be under these extreme conditions (necessity to measure ultra-fast low-amplitude signals) misleading.

One can see in the Fig.\,\ref{decay}a), that after the surface charging, the resulting compensating voltage may stabilize on various values (see ``After backward discharge" depicted values), depending on the stochastic nature of the process. 
The stochasticity is given by possible imperfections of the electrode surfaces, effectivity of the previous charge elimination process and the development of the forward and backward discharges themselves under the specific conditions. 
The measurement shown on Fig.\,\ref{decay} are single runs for given polarity and admixture. 
The error bars are given by the uncertainties of the laser parameters and the parabolic fitting of compensating voltage. 

In part b) of the Fig.\,\ref{decay}, the normalized representation of the compensating voltage/residual charge decay is shown. From these data it can be concluded that the admixture of the C$_{4}$F$_{7}$N leads to slower decay of the residual charge. 
The difference between the 10\,\% and 50\,\% admixture is not well defined. 
Additionally, the different polarity of the charging voltage pulse does not seem to influence the charge decay process consistently as well.

The presence of various additional chemical processes, including the decomposition of C$_{4}$F$_{7}$N in discharges of different nitrogen admixtures and polarities, is likely to lead to different dielectric surface modifications. 
For example, via thin layer or nanoparticles deposition onto the dielectrics (and the metallic electrode) by products of the reactions initiated by different plasma. 
This phenomenon may significantly influence not only the charge deposition, its magnitude, and spatial distribution but also the surface charge decay, because the modified surface
may have different conductivity or
may have modified or additional charge traps.

In Fig.\,\ref{decay}, a general trend is observed in pure nitrogen and nitrogen with C$_4$F$_7$N admixture, where higher initial compensating voltages are associated with higher concentrations of C$_4$F$_7$N. 
For charge deposition and decay in pure nitrogen and the 10\% C$_4$F$_7$N admixture approximately 20\%  of the initial charge is dissipated the first phase, which occurs within tens of minutes to a few hours.  
This is followed by a second phase, which begins roughly one hour after the start of the measurement, during which another 20\% of the charge decays over a time scale spanning tens of hours. 
In particular, the characteristic time for this second process is longer in mixtures containing C$_4$F$_7$N compared to pure nitrogen. 
A third slower decay is observed in the 50\% C$_4$F$_7$N admixture. 
This process operates on a timescale of hundreds of hours (or days) and is responsible for the dissipation of the remaining charge. 
Supporting evidence for this behavior is provided by measurements carried out over a one-week period for the 50\% admixture of C$_4$F$_7$N. 
Even after this extended time, the compensating voltage retained approximately 20\% of its initial value left after charge deposition, highlighting the significantly prolonged charge retention associated with higher C$_4$F$_7$N content.

\begin{table}
\centering
\renewcommand{\arraystretch}{1.4} 
\setlength{\tabcolsep}{14pt}     
\caption{Characteristic time constants for double (first four lines) and triple (last two lines) exponential model.}
\begin{tabular}{lccc}
\toprule
{Gas (polarity)} & {$\tau_\mathrm{A}$ [$10^{3} \cdot$ s]} & {$\tau_\mathrm{B}$ [$10^{5} \cdot$ s]} & {$\tau_\mathrm{C}$ [$10^{5} \cdot$ s]} \\ \midrule
N$_{2} (+)$ & 3.79 $\pm$ 0.61 & 1.38 $\pm$ 0.16 & - \\
N$_{2} (-)$ & 5.90 $\pm$ 1.17 & 0.88 $\pm$ 0.21 & - \\
N$_{2}$ + 10\% C$_{4}$F$_{7}$N ($+$) & 9.73 $\pm$ 2.02 & 3.91 $\pm$ 0.77 & - \\
N$_{2}$ + 10\% C$_{4}$F$_{7}$N ($-$) & 5.14 $\pm$ 0.74 & 6.10 $\pm$ 0.81  & - \\
N$_{2}$ + 50\% C$_{4}$F$_{7}$N ($+$) & 2.75 $\pm$ 0.55 & 0.45 $\pm$ 0.20 & 8.00 $\pm$ 2.86 \\
N$_{2}$ + 50\% C$_{4}$F$_{7}$N ($-$) & 1.53 $\pm$ 0.46 & 0.12 $\pm$ 0.02 & 3.96 $\pm$ 0.26 \\ \bottomrule
\label{times}
\end{tabular}
\end{table}

\begin{table}
\centering
\renewcommand{\arraystretch}{1.4} 
\setlength{\tabcolsep}{14pt}     
\caption{Weight factors for double (first four lines) and triple (last two lines) exponential model.}
\begin{tabular}{lccc}
\toprule
{Gas (polarity)} & {A [V]} & {B [V]} & {C [V]} \\ \midrule
N$_{2} (+)$ & 722 $\pm$ 23 & 741 $\pm$ 19 & - \\
N$_{2} (-)$ & 571 $\pm$ 28 & 635 $\pm$ 31 & - \\
N$_{2}$ + 10\% C$_{4}$F$_{7}$N ($+$) & 602 $\pm$ 27 & 1059 $\pm$ 29 & - \\
N$_{2}$ + 10\% C$_{4}$F$_{7}$N ($-$) & 380 $\pm$ 11 & 1779 $\pm$ 10  & - \\
N$_{2}$ + 50\% C$_{4}$F$_{7}$N ($+$) & 399 $\pm$ 20 & 559 $\pm$ 71 & 2121 $\pm$ 87 \\
N$_{2}$ + 50\% C$_{4}$F$_{7}$N ($-$) & 424 $\pm$ 33 & 789 $\pm$ 34 & 2177 $\pm$ 25 \\ \bottomrule
\label{weights}
\end{tabular}
\end{table}

 The characteristic time constants with their weight factors for the data presented in Fig.\,\ref{decay}, as represented by equation\,\ref{exp}, are shown in table\,\ref{times} and table\,\ref{weights}, respectively. 
 The time constants $\tau_\mathrm{A}$ and $\tau_\mathrm{B}$ have clearly different scales, $\tau_\mathrm{A}$ in thousands and $\tau_\mathrm{B}$ in hundreds of thousands of seconds.
The time constants for both polarities of the 10\,\% admixture of C$_4$F$_7$N deviate also from possible linear trend between the other two mixtures (pure nitrogen and 50\,\% C$_4$F$_7$N). 
Apparently, a small 10\,\% admixture of C$_4$F$_7$N changes the physics on the dielectric surface differently and cannot be seen simply as a step in between the pure nitrogen and 50\,\% C$_4$F$_7$N admixture. The weight factors in table\,\ref{weights} mirror the same tendency.

\begin{figure}[!h]
    \centering
    a)
\includegraphics[width=.45\textwidth]{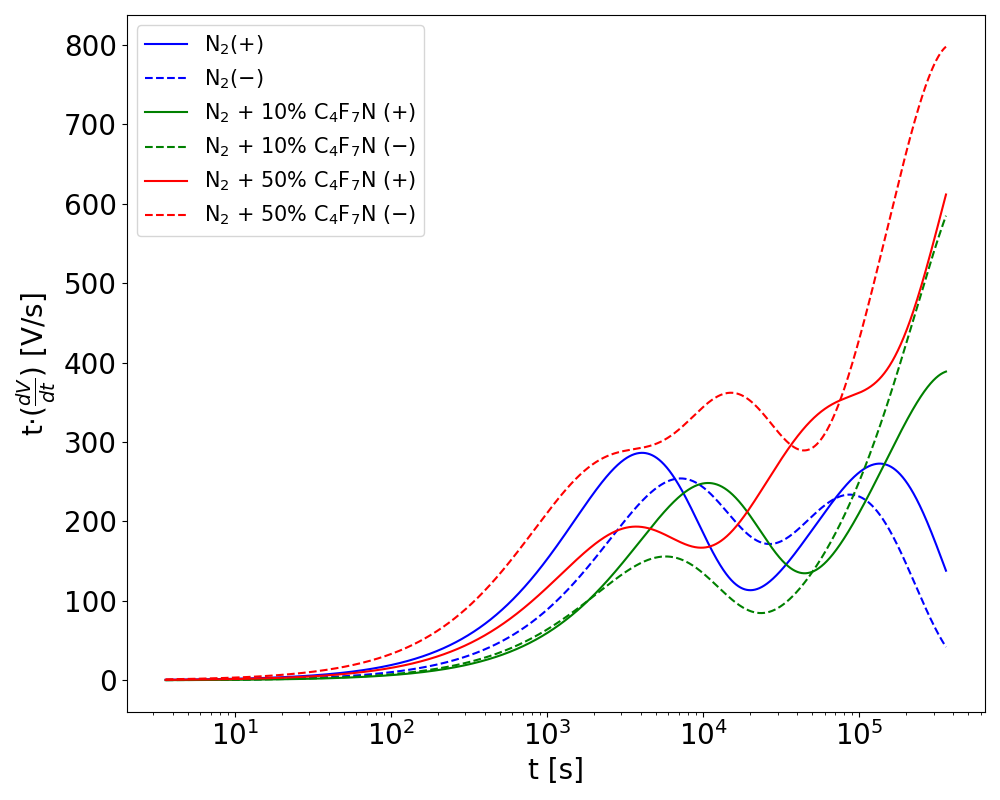}
b)
\includegraphics[width=.46\textwidth]{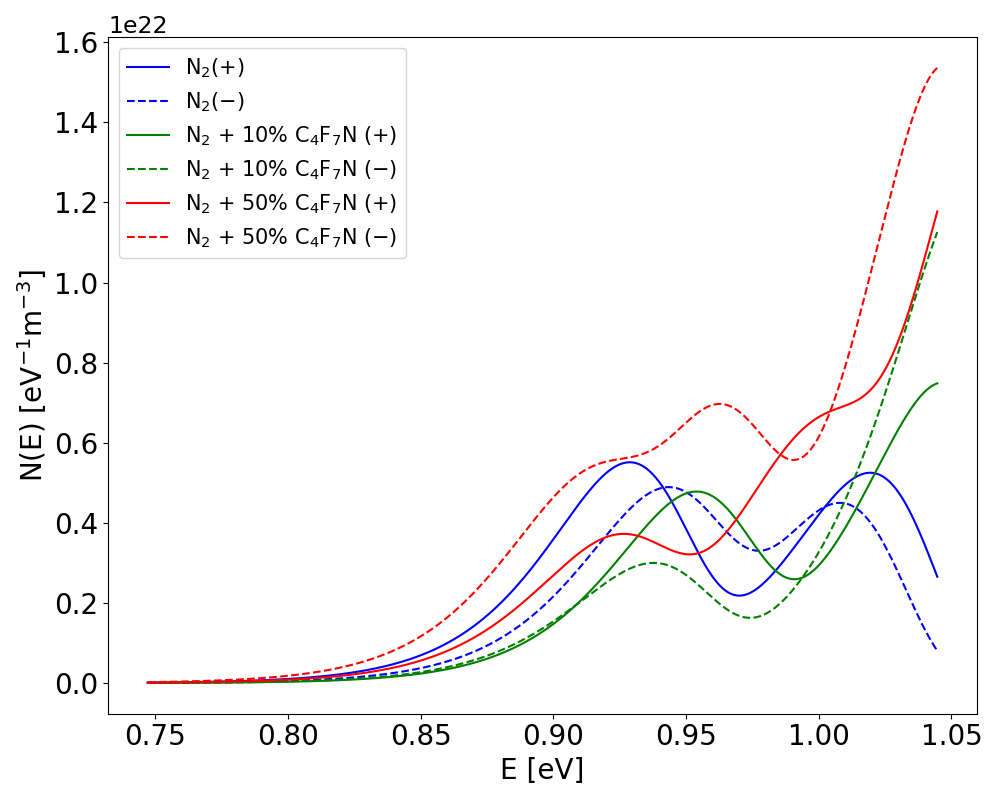}
\caption{Charge decay curves as given by compensating voltage measurements shown in Fig.\,\ref{decay} recomputed using equation (\ref{trap1}) are shown in part a). In part b), the equations (\ref{trap2}) and (\ref{trap3}) are used.}
\label{traps}
\end{figure}

In order to analyze the surface charge decay curves, the isothermal model for surface charge decay controlled by charge detrapping was utilized. 
The model is based on the assumption that as time elapses (for given temperature), the charges trapped in the energetic states are progressively released and that the charges emitted at a given time correspond to a preferential trap depth or the so-called demarcation energy $E$ (i.e. the latter emitted charge, the higher the energy of the trap)\cite{molinie2024,teyssedre2021}. 
The trap distribution and their relative densities can be represented using following equations:

\begin{equation}
\label{trap1}
t\,\frac{dV_{s}}{dt}\;\propto\; k_{\mathrm{B}}\,T\,N\!\bigl(E_{t}(t)\bigr)
\end{equation}

\begin{equation}
\label{trap2}
E_{t} \;=\; k_{\mathrm{B}}\,T\,\ln\!\bigl(\nu\,t\bigr),
\end{equation}

\noindent
i.e. that the term $t\,\frac{dV_s}{dt}$ is proportional to the charge trap density whereas the term $\ln\!\bigl(\nu\,t\bigr)$ is proportional to energy of the trap (the energy difference between the trap energy and the conductive band). 
The results in Fig.\,\ref{decay}, rescaled using the equation (\ref{trap1}), are shown in Fig.\,\ref{traps}a).  
The eq.\,(\ref{trap1}) can be also exactly given in following form (as developed by \cite{simmons1973}):

\begin{equation}
\label{trap3}
N\!\bigl(E_{t}\bigr)\;=\;
\frac{\varepsilon_{0}\,\varepsilon_{r}\,t}{k_B\,T\,f_{0}\!\bigl(E_{t}\bigr)\,\delta L\,q}\;
\frac{dV_{s}}{dt}\,,
\end{equation}

\noindent
where in the above mentioned equations the $V_s$ represents the measured voltage (in our case the compensating voltage), $t$ is time, $k_B$ Boltzmann constant, $N(E_t(t))$ is energy/time dependent charge trap density, $T$ is the temperature (300\,K in our case), $E_t$ trap energy, $\nu$ is attempt-to escape frequency of the trapped charge (10$^{12}$\,s$^{-1}$ in our case \cite{zhang2018}), $\varepsilon_0$ is the vacuum permittivity, $\varepsilon_r$ is the relative permittivity of the alumina (equals 9 for our case), $f(E_t)$ is the rate of initial occupancy of traps (for simplicity equals 1 as in \cite{zhang2018} which means that all traps were fully occupied), $L$ is the sample thickness (equals 1\,mm for our case) and $\delta$ is the thickness of top charge layer (2\,$\mu$m, see \cite{zhang2018}). 

The charge decay results recomputed using eq.\,(\ref{trap2}) and (\ref{trap3}) are shown in Fig.\,\ref{traps}b). 
The trap distribution for charge decay after the discharge in nitrogen atmosphere is consistent with results found in literature \cite{zhang2018,svandova2024,wang2019} for similar alumina samples (96\,\% purity) which were obtained via well established experimental procedure using electrostatic voltmeter and Kelvin probes. 
Particularly, two peaks are visible in Fig.\,\ref{traps}b) as reported previously in the mentioned literature, see the blue curves for both polarities. 
One peak lies at approximatelly 1.01\,eV and the other at 0.93\,eV. 
Similar trap energies for charges deposited using a barrier discharge were found also using thermoluminescence and optically stimulated luminescence in \cite{ambrico2009}.

For the charge decays deposited in atmospheres with C$_4$F$_7$N admixtures, the trap distribution is different. For 10\,\% C$_4$F$_7$N admixture it seems that, consistently for both polarities, the traps shifted towards higher energies, having the maxima around 1.05\,eV at the border of the measurement interval. It is depicted by the green curves. 
This limit is in the isothermal charge decay model given by the longest time of surface charge decay measurements, as it follows from equation\,(\ref{trap2}). 
For 50\,\% C$_4$F$_7$N admixture, the situation changes again, see the red curves in Fig.\,\ref{traps}b). Three maxima are visible from the model (given by the necessary three exponential decay functions in the fit) showing a possible splitting of the peak near 0.93\,eV. 
The shift of known traps or possibly creation of new traps in the case of C$_4$F$_7$N admixtures is connected to the surface modification by plasma in carbon and fluorine containing atmospheres. Apparently, under given conditions it leads to establishing of deeper traps and therefore more long-lasting charge residual at the dielectric surface. It is the exact opposite as expected, see for example \cite{wang2019}.

\begin{figure}[!h]
    \centering
    a)
\includegraphics[width=0.452\textwidth]{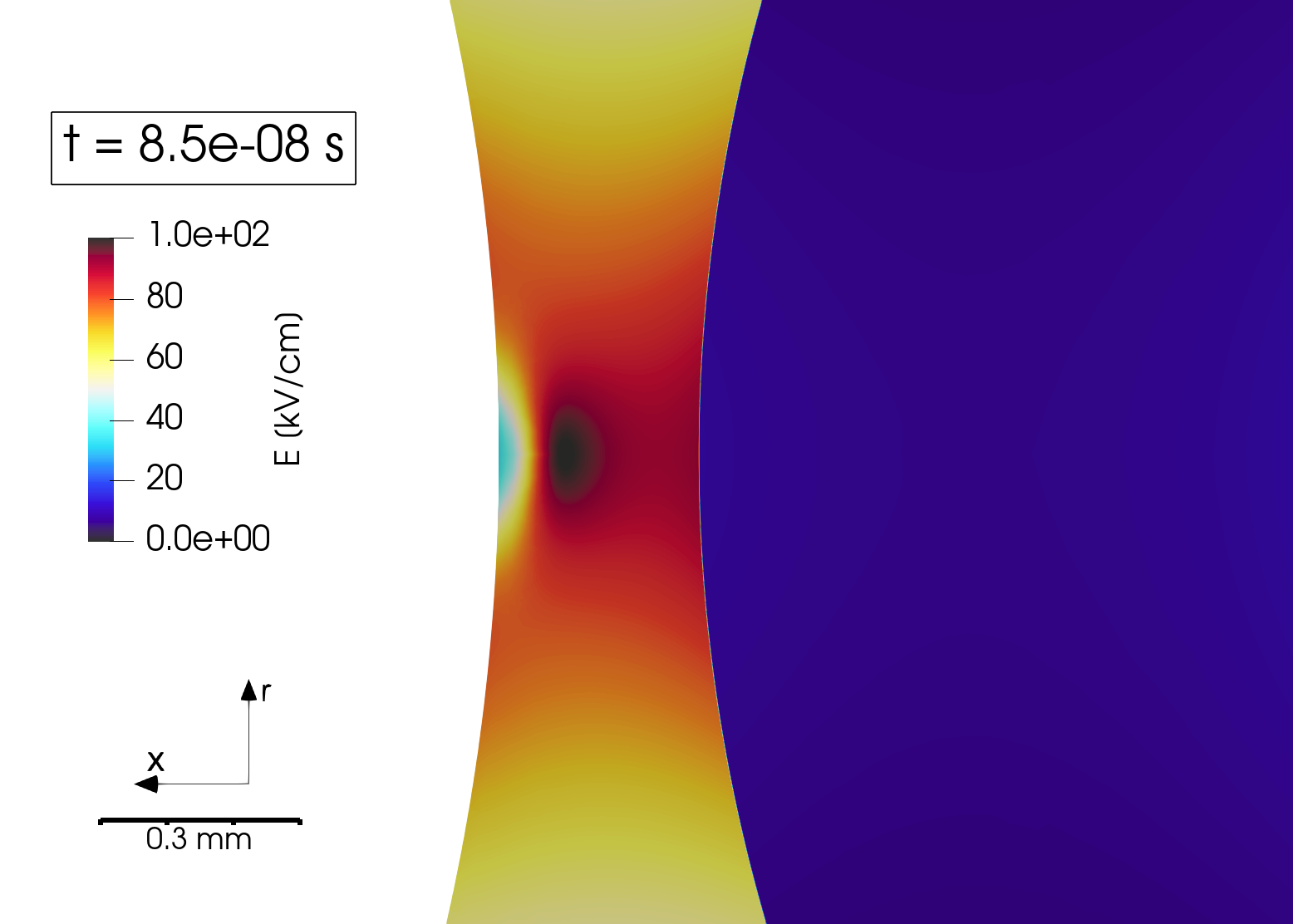}
b)
\includegraphics[width=0.452\textwidth]{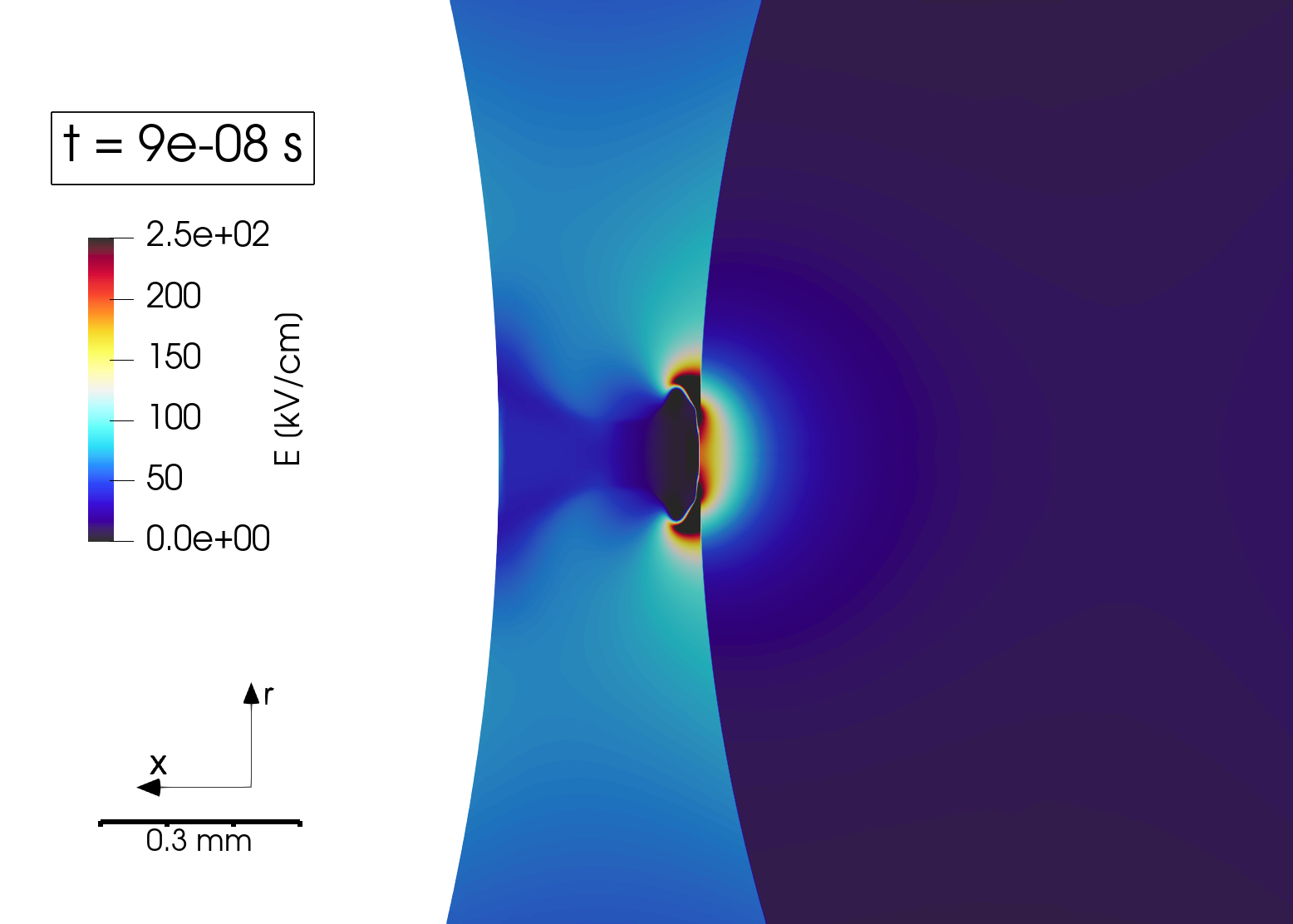}
c)
\includegraphics[width=0.452\textwidth]{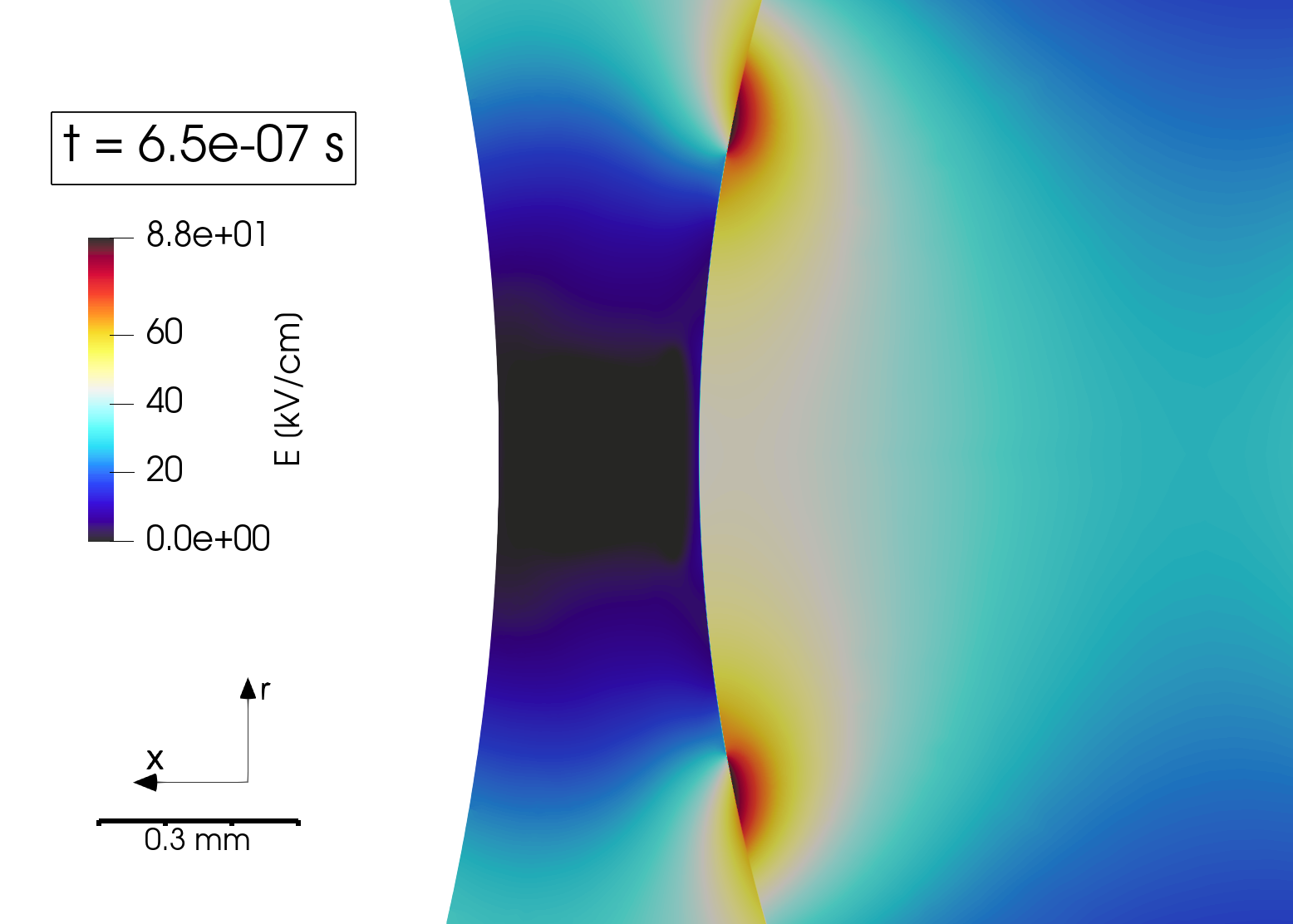}
d)
\includegraphics[width=0.452\textwidth]{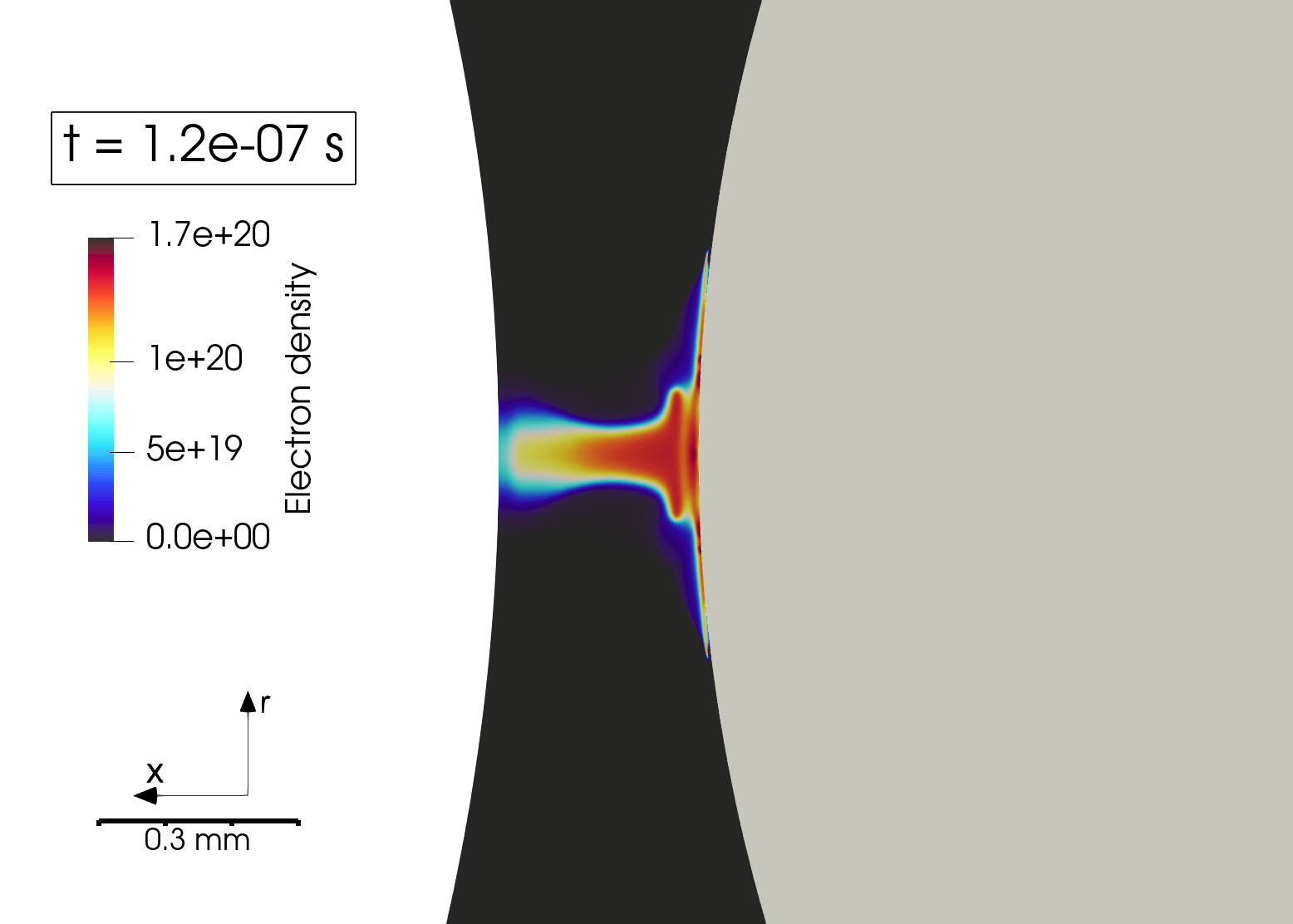}
\caption{Results of the numerical modeling in nitrogen for positive applied voltage polarity on the powered electrode. Magnitude of the electric field of the simulated streamer during streamer initiation, part a). Its magnitude during impact on the solid dielectric, part b), and the electric field magnitude of the simulated streamer at the last simulated time, part c). In part d), the electron density of the simulated streamer at $t=120$\,ns is shown.}
\label{fig:E_initiation}
\end{figure}

It is important to note that the exact analysis of the surface charge decay curves is complex \cite{molinie2024,teyssedre2021}. 
The presented results should be therefore understood as initial estimates based on the novel in-operando EFISH method. 
Typically, the surface conduction needs to be identified first in the decay. 
The results presented in \cite{zhang2018} show that at the alumina dielectrics charged in nitrogen containing gas the surface conduction is not present. 
The EFISH method presented here is utilized for electric field measurement in one coordinate only and spatially resolved measurements for surface conductivity clarification are intended in the future. 
Further, the high electric field during the charge deposition is important. In barrier discharges, its value can reach up to few hundreds of kV/cm, see Fig.\,\ref{fig:E_initiation} or \cite{jansky2021}. 
Given the agreement of the presented results with the literature for charge decay on alumina surfaces charged in nitrogen containing discharges, we believe that the new method showed its potential.

\subsection{Streamer simulation results}
In order to gain further insight into the discharge dynamics and to confront the experimental data with theory, numerical modeling for pure nitrogen has been done in axi-symmetric geometry of studied system, as depicted in Fig.\,\ref{komora}b).
The applied voltage was set to $V_0 = 3.5$\,kV, as this value was the lowest breakdown voltage measured in the experiments for pure nitrogen, positive polarity.
The relative permittivity of the solid dielectric was set to $\varepsilon_\mathrm{d} = 9$, as described in the section on the numerical modeling.

\begin{figure}[!h]
    \centering
\includegraphics[width=0.5\textwidth]{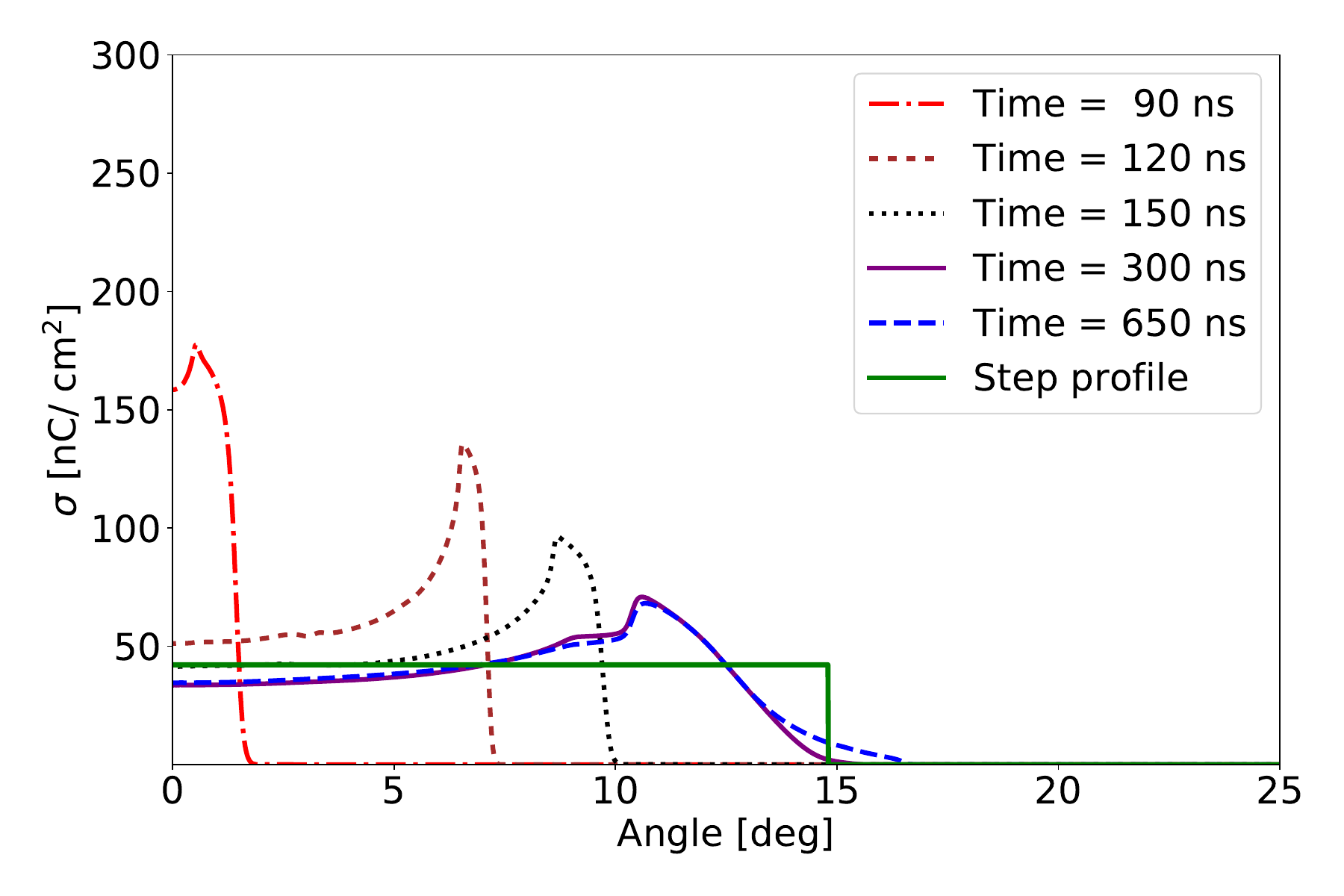}
\caption{Surface charge density on the dielectric in the streamer simulation at various times as a function of polar angle measured from the axis, see the next figure. Also shows a step profile of surface charge with the same compensating voltage and same surface charge the one from streamer simulation at $t=650$\,ns.}
\label{fig:sigma}
\end{figure}

The evolution of the positive streamer in pure nitrogen is shown in Fig.\,\ref{fig:E_initiation}.
The potential on the left (bare) electrode was set to $3.5$\,kV, the electrode covered by dielectrics (right) is grounded. Note that only the dielectric barrier is visible in the figure, while the interface to the electrode is located at the right-hand edge of each sub-figure. 
The streamer initiation occurs at around $60$\,ns after simulation initiation and few nanoseconds later the streamer impacts and deposits the dielectric surface.
The values of the electron density (up to 10$^{20}$\,m$^{3}$) as well as of the electric field (up to 250\,kV/cm in the dielectric surface vicinity) are consistent with the results presented in \cite{papageorghiou2009}. 
The simulation was stopped at $650$\,ns as the rate of change in the surface charge density was very low. In Fig.\,\ref{fig:sigma} presented change in the surface charge density between $300$\,ns and $600$\,ns is negligible.

The total surface charge after integration of the surface charge density profile in Fig.\,\ref{fig:sigma} is $q = R^2 \int_0^{2 \pi} \mathrm{d} \phi \int_0^{\pi} \mathrm{d} \theta ~\sigma(\theta) \sin \theta = 0.55$\,nC, where $R = 2.5$\,mm is the radius of the dielectric and $\phi$ and $\theta$ are the azimuthal and polar angles, respectively. 
The streamer simulation results in a transferred charge of 0.55\,nC for one positive discharge (i.e. the first forward discharge in positive polarity). 
The transferred charge under the same conditions in experiment is $0.43$\,nC (see table\,\ref{forwardpulse}). 
Thus, the total charge transferred to the dielectric is overestimated in streamer simulation by $0.12$\,nC compared to the experimentally measured average value 
obtained from electric current integration. 
Note that the simulation represents only a forward discharge to uncharged dielectrics, without backward discharge as it is the case in the experiment. 
The mentioned difference of 0.12\,nC for the forward discharge (between experiment and simulation) could be due to a number of factors. 
The axi-symmetric approach, chosen here for computational efficiency, cannot account for the streamer branching on the surface of the dielectrics. 
Also, the close proximity of the electrode and dielectrics to the plasma region may result in uncertainties, as the local field approximation may be insufficient.
Using a model with mean electron energy equation would likely result in additional cooling of electrons near the boundaries and, therefore, slower streamer propagation and lower ionization \cite{Viegas2022}. 
The charge estimation based on the electric measurements may also contain some uncertainty due to the residual charges from previous discharges (see the discussion regarding the table\,\ref{chargebalance}).

The simulation also predicts the spread of the surface charge over the dielectrics. 
At the beginning of the charging process (e.g. at 90\,ns, see in Fig.\,\ref{fig:sigma}), the charge density is 
about $150$\,nC/cm$^{-2}$ in a very narrow region close to the axis.
Later at the end (650\,ns), the dielectric surface is charged to angle of about $15^\circ$ by surface charge density of approximately $35$\,nC/cm$^{-2}$ in the region closer to the axis of the rotational symmetry. 
It has nevertheless a maximum, almost at its edge, with charge density of 
about $70$\,nC/cm$^{-2}$. 
In the further analysis, the use of the step profile approximation of the surface charge density distribution (see the solid green line in Fig.\,\ref{fig:sigma}) will be utilized and justified. 

\subsection{Linking compensating voltages from EFISH to surface charge density}\label{sec:modelresults}

\begin{figure}[!h]
    \centering
\includegraphics[width=0.3\textwidth]{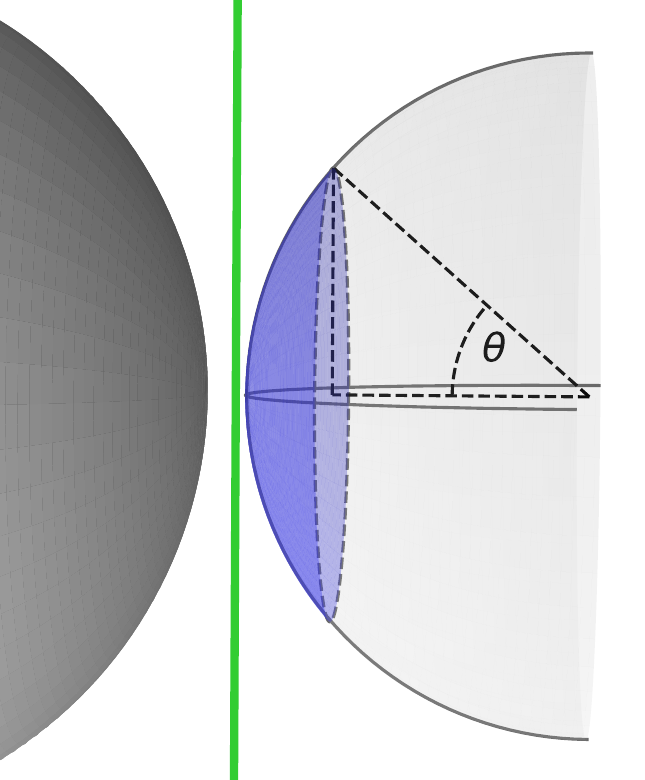}
\caption{Setup for electrostatic calculations resulting in electric field distribution in quasi-3D, i.e. in 2D axi-symmetric system. The bare electrode on the left, the dielectric-covered electrode on the right. The surface charge is set to constant value in the blue region characterized by the polar angle $\theta$. The laser path is shown as a green line.}
\label{fig:lim_angle}
\end{figure}

Additional electrostatic calculations were performed to provide interpretation of 
EFISH measurements and to enhance surface charge analysis.
 These computations determined the two-dimensional axi-symmetric spatial distribution of the electric field in the system for given deposited charge and its density on the surface, i.e. its spread on the dielectrics, and at given applied voltage. 
 The electrostatic calculations were done by solving the Poisson's equation, in the same way as for the streamer simulations, using finite volume method and an axi-symmetric setup.
For simplicity, the surface charge placed on the dielectric was modeled as a step function (see the green line in Fig.\,\ref{fig:sigma}) along the polar angle, i.e. before a certain limiting angle the surface charge density is constant and after it equals to zero (see also the 3D representation of the surface charge spread in Fig.\,\ref{fig:lim_angle}).

The electrostatic calculations allow us to connect the measured compensating voltages from the EFISH with electrical diagnostic techniques and to access the surface charge density on the dielectric as indicated by streamer simulations, see Fig.\,\ref{fig:sigma}.
The step function of the surface charge density profile is given by two parameters: the total charge and the limiting angle.
Since the modeled electric field is axi-symmetric, the EFISH signal intensity measured along the laser path (green line in Fig.\,\ref{fig:lim_angle}), given by equation\,\ref{eq_efish}, can be simplified to:
\begin{equation}
    I_\mathrm{EFISH} \propto \left( \int_0^\infty E_x(r)  \frac{z_R \cos(\Delta k r) + r \sin(\Delta k r)}{z_R^2 + r^2} \mathrm{d}r \right)^2
    \equiv \left( \int_0^\infty E_x(r) w(r) \right)^2
    \label{eq:efish_model}
\end{equation}
where $r$ is the distance direction from the axis of rotational symmetry, $E_x$ is the axial component of the electric field, $w$ is the weight that depends on the laser setup.
The wave vector mismatch was calculated based on the refractive index for nitrogen \cite{Brzsnyi2008}, $\Delta k = 4 \pi \left(\frac{n(\lambda_0)}{\lambda_0} - \frac{n(\lambda_1)}{\lambda_1}\right) = -49$\,m$^{-1}$, where $\lambda_0 = 1064$\,nm and $\lambda_1 = 532$\,nm is the fundamental and second harmonic wavelength, respectively.
The Rayleigh length for the experimental optical system was estimated at $z_R = (8 \pm 4) \cdot 10^{-4}$\,m. 

To find the compensating voltage for a given surface charge profile, the electrostatic calculations were run iteratively. For various total charges on the dielectrics and various limiting angles, i.e. for different surface charge densities given the step function, the applied voltage between the electrodes was varied until the minimum of EFISH signal intensity, given by equation\,\ref{eq:efish_model}, was found.
The integral was evaluated numerically using the trapezoidal rule with the upper integration bound set to $13$\,mm instead of infinity. 
It was verified that, for the parameters considered in this study, the contribution to the integral at the upper bound ($z = 13$\,mm) is four orders of magnitude lower than in the region near the axis. 
Thus, negligible contributions from outside the domain were ignored.

\begin{figure}[!h]
    \centering
\includegraphics[width=0.5\textwidth]{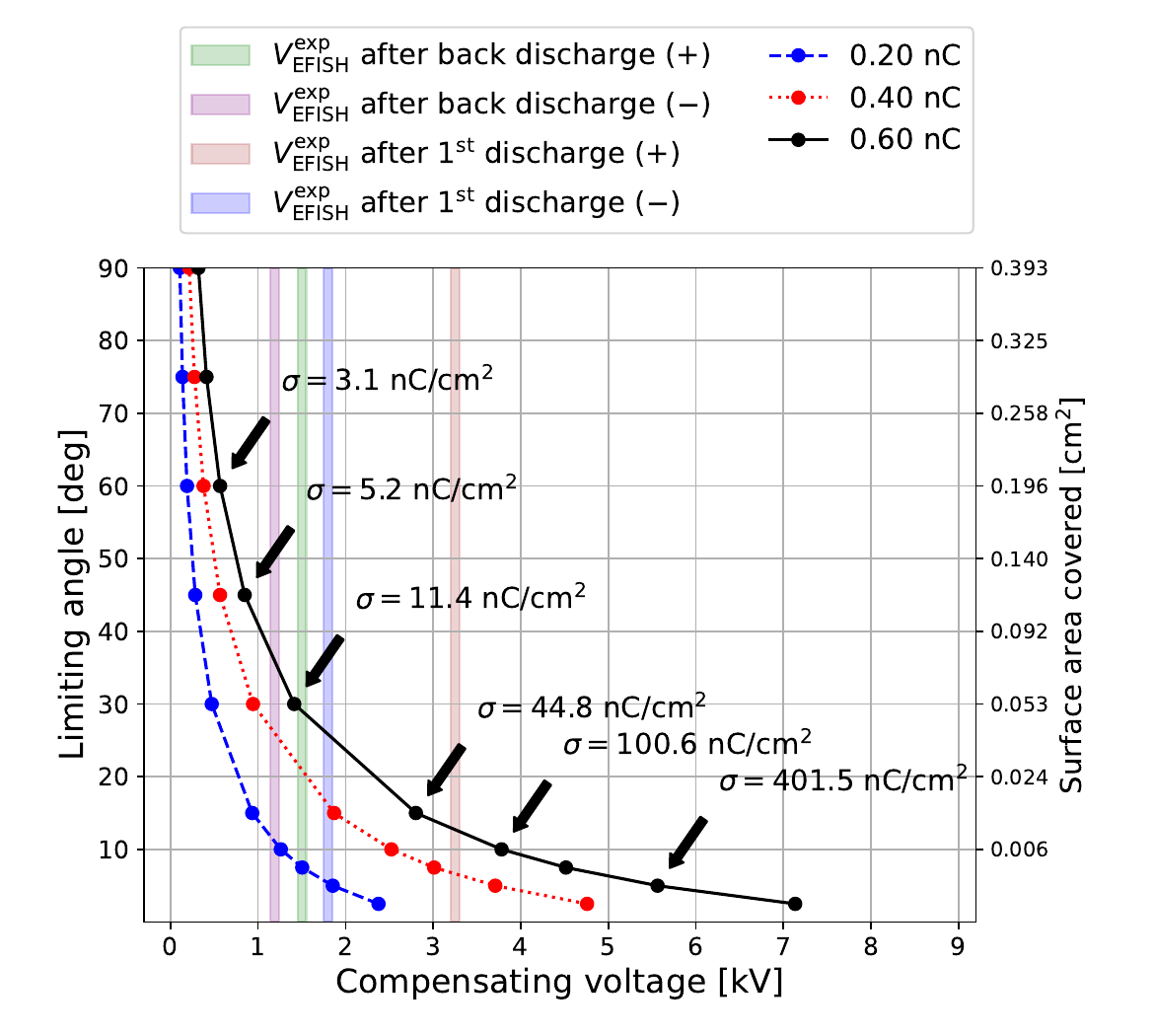}
\caption{Limiting angle and surface area of homogeneous surface charge covering symmetrically the cone of the solid dielectric for various total charges as functions of voltage between the electrodes that produces the minimal EFISH signal. This voltage has been determined with electrostatic simulations. 
The colored columns are experimental data from table \ref{tab:voltage_comp}, for compensating voltage after the forward and backward discharges, and after a single forward discharge as shown in Fig.\,\ref{f_efishparaboly}. }
\label{fig:comp_voltage}
\end{figure}

The results of the electrostatic calculations are shown in Fig.\,\ref{fig:comp_voltage} for total surface charges of $0.2$, $0.4$ and $0.6$\,nC, in the range of the experimentally obtained values.
Apparently, the compensating voltage becomes very sensitive to the value of the limiting angle below $30^\circ$.
These results clearly show what is the spread of charge distribution (given by the limiting angle or covered surface area). The spread can be then determined from given transferred charge (experimentally accessible by electrical current measurements) and compensating voltage value (accessible by EFISH measurements) as parametrized in Fig.\,\ref{fig:comp_voltage}. 

\begin{figure}[!h]
    \centering
\includegraphics[width=0.5\textwidth]{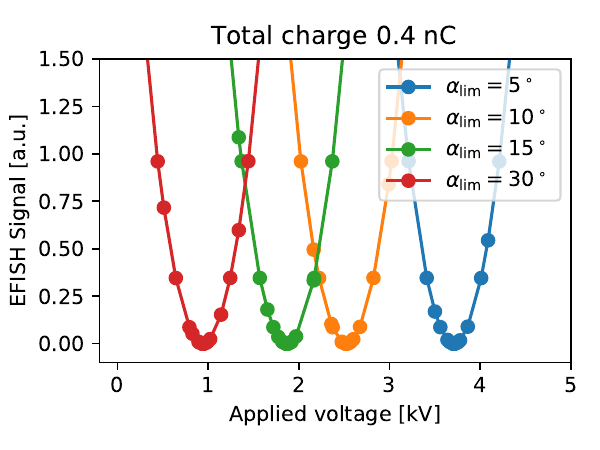}
\caption{Simulated EFISH signal for total charge on the dielectric of $q = 0.4$\,nC  using a step profile of surface charge density and various limiting angles.}
\label{fig:sim_parabolas}
\end{figure}

In Fig.\,\ref{fig:sim_parabolas}, computed EFISH signal (see equation (\ref{eq:efish_model})), for a simulated electric field in the gap, is presented for several spreading angles using a fixed total surface charge and for different applied voltage values. 
Clearly, for constant amount of deposited charge of 0.4\,nC, the system manifests following feature: the smaller the angle (i.e. the more is the charge confined to a smaller area) the bigger is the electric field at probed coordinate and the higher goes the compensating voltage.

\begin{figure}[!h]
    \centering
\includegraphics[width=0.5\textwidth]{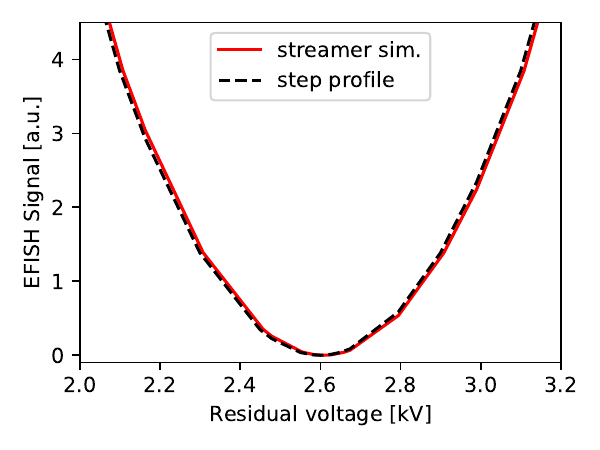}
\caption{Comparison of the calculated EFISH signal from the surface charge predicted by the streamer simulation and the surface charge with step profile with the minimum at the same compensating voltage as in the streamer simulation.}
\label{fig:efish_sim}
\end{figure}

The justification of the utilized step profile approximation was made in the following way. 
The compensating voltage for the surface charge of 0.55\,nC, resulting from the streamer simulation for $t = 650$\,ns (see the blue dashed line in Fig.\,\ref{fig:sigma}), was found equal to $2.6$\,kV. 
Furthermore, using the compensating voltage of 2.6\,kV and the total surface charge of 0.55\,nC from streamer simulation, we can calculate the equivalent limiting angle for the step profile used in the electrostatic calculations. 
It results in limiting angle of $\theta_\mathrm{lim}(q = 0.55~\mathrm{nC}, V_\mathrm{C} = 2.6~\mathrm{kV}) = 14.8^\circ$ and corresponds to the surface charge density of $42.2$\,nC/cm$^2$ - giving thus the simplified step profile function. 
Considering this simplified step profile instead of the original charge density profile from streamer simulation makes almost no difference to the integration of equation\,\ref{eq:efish_model}.
It can be seen from Fig.\,\ref{fig:efish_sim}, where the integral for the two cases is compared as a function of the applied voltage and shows virtually no difference. 
The axial electric field $E_x$, used for the calculation of the EFISH signal (when the compensating voltage is applied) is very similar, both for the step profile charge density and for the original streamer simulation profile. 
Most of the deviation occurs near the axis of rotational symmetry (cosine similarity, a measure of similarity, is of $>99$\% in the integrated region).

\begin{figure}[!h]
    \centering
    a)
\includegraphics[width=0.45\textwidth]{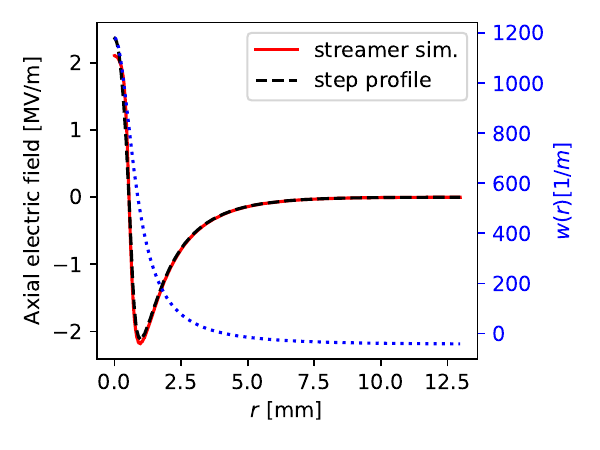}
b)
\includegraphics[width=0.45\textwidth]{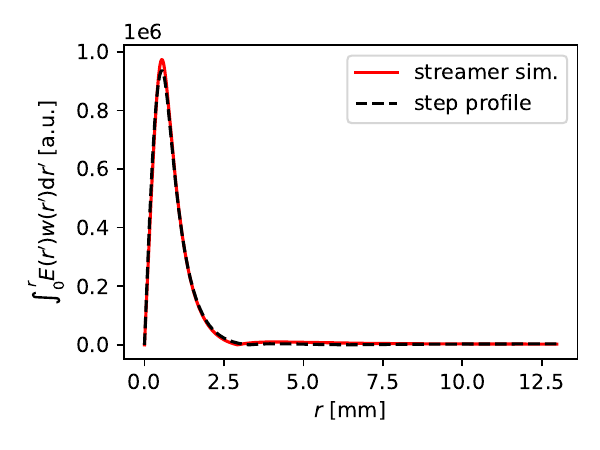}
\caption{In part a), a comparison is shown of the axial component of the electric field $E_x$ for the step profile surface charge and the surface charge taken from the streamer simulation at $t=650$\,ns. The integral weight $w$ used in EFISH measurements for the parameters of the laser used in this study is presented, too. In part b), the cumulative sum of the EFISH integrant for the streamer simulation surface charge and the surface charge with the step profile is depicted. The step profile surface charge value was chosen to match the total amount of charge on the surface in the streamer simulation. The limiting angle was chosen such both profiles produce the same EFISH intensity.}
\label{fig:sim_signal}
\end{figure}

\begin{figure}[!h]
    \centering
\includegraphics[width=0.5\textwidth]{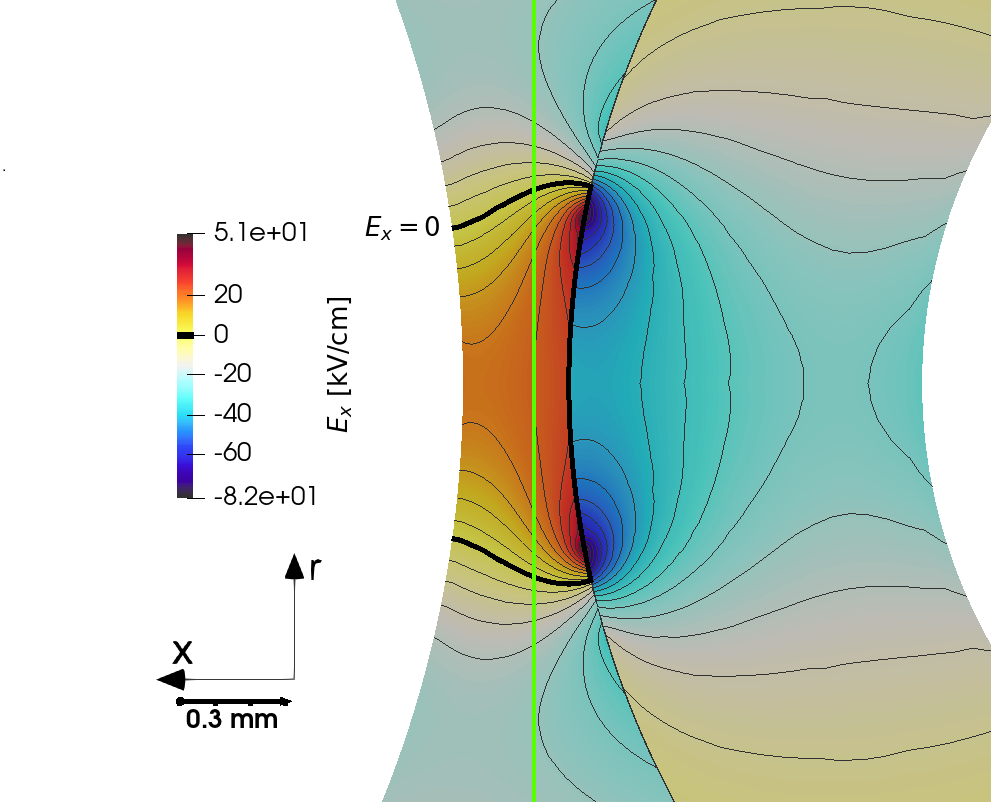}
\caption{Axial component ($x$-axis towards metallic electrode) of the electric field with the surface charge on the dielectric taken from streamer simulation ($t = 650$\,ns) and with applied compensating voltage $V = 2.6$\,kV. Note that the $E_x$ = 0\,kV/cm is denoted by bold curve.}
\label{fig:E_compensating}
\end{figure}

It is important to note that although the voltage applied in the streamer simulation is only $3.5$\,kV, the  compensating voltage from electrostatic calculations is much lower, only $2.6$\,kV. 
This difference highlights the complex meaning of the compensating voltage. 
As it is not a simple measure of the zero (eliminated) electric field in the gap, but certain value of external applied voltage for which the integral in equation\,\ref{eq:efish_model} is minimized. 
Integral in equation\,\ref{eq:efish_model} consist of two components, $E_x$ and $w$, integrated over the interaction path of the laser in radial distance $r$ (along the $z$-axis). 
The axial electric field component distribution $E_x$ results from contribution of applied voltage on the high-voltage metal electrode and the electric field originating from the surface charge. The $w$ component results from the laser parameters, as described above. The development of these two functions is visualized in Fig.\,\ref{fig:sim_signal}a).

The distribution of the electric field of the axial component, obtained for the system when the integral in equation\,\ref{eq:efish_model} is minimal, is shown in Fig.\,\ref{fig:E_compensating} and the cumulative sum along the line of integration in Fig.\,\ref{fig:sim_signal}b).
The axial direction ($x$-axis) is oriented from the solid dielectrics towards the electrode.
With a positive surface charge the axial electric field value is positive on the axis of rotational symmetry for small values of applied voltage.
However, when the surface charge is limited to a conical section of the dielectrics, and when a positive voltage is applied, the axial electric field will be negative from a certain distance from the axis, see Fig.\,\ref{fig:sim_signal}a) and \ref{fig:E_compensating}. 
Therefore, the integral\,\ref{eq:efish_model} will be minimal when contributions to the axial electric field from the surface charge (positive electric field) balance out the contributions from the applied voltage (negative electric field).
Since the weight in the integral, given by the optical parameters, is the highest around the axis and quickly decays, as does the electric field magnitude, the main contributions to the integral are in the gap region (close to the interelectrode axis) and the electric field further away has negligible influence on the result. 

It is apparent quantitatively from the above-mentioned analysis that the compensating voltage should not be interpreted as the voltage that minimizes the maximal electric field in the 
gap. 
It is merely the voltage that minimizes the integral\,\ref{eq:efish_model} for given electric field distribution.
The applied voltage value that produces zero electric field on the axis of rotation in the gap may be quite different in this geometry, especially with concentrated surface charge as the axial electric field is reversed closer to the focal point of the laser.

Coming back to the comparison with experiment in Fig.\,\ref{fig:comp_voltage}, the experimental values of compensating voltage are shown with colored columns and taken from table \ref{tab:voltage_comp} and Fig.\,\ref{f_efishparaboly} for the case of nitrogen.
According to current measurements, the charge remaining on the dielectrics after forward and backward discharges had average values of 0.14\,nC and 0.40\,nC for the positive and negative polarity, respectively (see table \ref{chargebalance}).
These charge values, together with the compensating voltages measured by EFISH of 1500\,V for positive and 1190\,V for negative polarity (table \ref{tab:voltage_comp}), give an estimated spread of surface charge of $<$10$^\circ$ for the positive polarity and $\approx$ 20$^\circ$ for the negative polarity (see Fig.\,\ref{fig:comp_voltage}).

Roughly the same angles are obtained when looking at the values measured after the forward discharge only:
The charge $Q_\mathrm{F}$ after the forward discharge had average values 0.43\,nC and 0.52\,nC for the positive and negative polarity, respectively (see table \ref{forwardpulse}), while the compensating voltage was $\approx$3.25\,kV and $\approx$1.8\,kV for the positive and negative polarity, respectively (Fig.\,\ref{decay}).
This again gives angles of $<$10$^\circ$ for the positive polarity and $\approx$20-30$^\circ$ for the negative polarity. 
The total charge on the dielectric obtained from the streamer simulation (0.55\,nC) is higher than the experimentally measured value (0.43\,nC), and so is the limiting angle: 15$^\circ$ from the simulation and $<$10$^\circ$ calculated from the experimental values.
Under given conditions, this is relatively good agreement. 
The discrepancy may be due to approximations of the fluid model and the boundary conditions used at the dielectric interface. 
Furthermore, the uncertainty in the focal position of the laser has been considered as a potential cause for an error. 
Nevertheless, a shift in the focal position of the laser during the measurement would not explain this discrepancy.

\subsection{Linking electrically measured effective compensating voltages to surface charge density}

Similarly as with the EFISH measurements, the voltage values measured with electrical diagnostics, presented in table\,\ref{tab:voltage_comp}, may also be linked with the surface charge using electrostatic calculations.
Here, we focus again on the pure nitrogen measurements as in the previous case for EFISH.
To find the breakdown voltage increase, $V_\mathrm{B}^\ast - V_\mathrm{B}$, i.e. the electrically determined effective compensating voltage, we assume that the breakdown with the surface charge happens at the same electric field (breakdown field in the gas gap) as in the case without any surface charge.
For the theoretical analysis here, the breakdown voltage is determined using electrostatic simulations as applied voltage value that under given conditions generates the breakdown field at the center of the gap, at $x=0.15$\,mm.

\begin{figure}[!h]
    \centering
\includegraphics[width=0.5\textwidth]{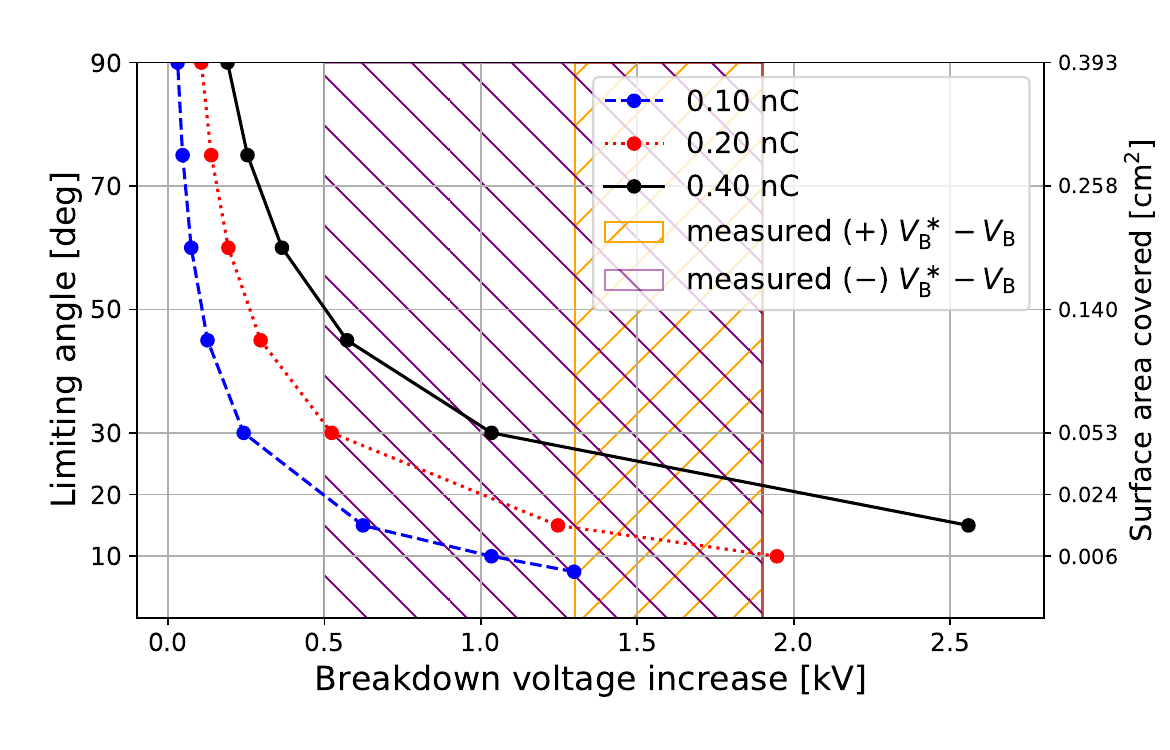}
\caption{Limiting angle and surface area for  homogeneous surface charge covering symmetrically the cone of the solid dielectric for various total charges as functions of the additional applied voltage needed for the breakdown field (i.e. the breakdown voltage increase $V_\mathrm{B}^\ast - V_\mathrm{B}$) for pure nitrogen with positive polarity (see first column of the first row of table\,\ref{tab:voltage_comp}). Experimental measurements shown with hatched areas.}
\label{fig:bd_voltage_calc}
\end{figure}

The measured breakdown voltage with eliminated surface charge for positive polarity in pure nitrogen is $V_\mathrm{B} = 4.5$\,kV (see table\,\ref{tab:voltage_break}). 
For such voltage, the electrostatic calculations result in electric field in the middle of the gap with value of $108$\,kV/cm.
For the negative polarity 
this electric field is lower as the breakdown voltage is lower (3.7\,kV), and 
the electrostatic calculations results in the field of $89.2$\,kV/cm.
This is in excellent agreement with the estimates using the planar dielectric approximation presented in previous section, see equation\,\ref{efield} and table \ref{tab:field}.

The results presented in Fig.\,\ref{fig:bd_voltage_calc} show the breakdown voltage increase values with deposited surface charge for its various limiting angles and total charge values.
The limiting angle, as a function of breakdown voltage increase $V^\ast_\mathrm{B} - V_\mathrm{B}$, is identical for positive and negative polarity and starts to deviate below the limiting angle of $10^\circ$. 
For positive polarity, the breakdown voltage increase $V^\ast_\mathrm{B} - V_\mathrm{B}$ is equal (1600$\pm$300)\,V (see the table \ref{tab:voltage_break}) which for surface charge of (0.14$\pm$0.04)\,nC (see the table \ref{chargebalance}) results in limiting angle for surface charge spread of about $10^\circ$ and below. 
For negative polarity (1200$\pm$700)\,V and (0.40$\pm$0.05)\,nC it results in limiting angle for surface charge spread of about $30^\circ$ and below.

As is apparent from Figs.\,\ref{fig:comp_voltage} and \ref{fig:bd_voltage_calc} the compensating voltage (EFISH) and the voltage increase from electrical diagnostics are quite sensitive at the low values of limiting angles.
Given this and the statistical uncertainties in the estimation of optical parameters of the EFISH laser, charge measurements and approximation in our streamer model we cannot predict the precise profile of the surface charge with certainty.
However, in conclusion, EFISH and electric diagnostic measurements and simulation results for pure nitrogen and positive polarity are consistent in predicting that the area to which the surface charge spreads is limited to a conical section with limiting angle below $15^\circ$. 
For negative polarity, both EFISH and diagnostic measurements predict higher limiting angles as seen from Figs.\,\ref{fig:comp_voltage} and \ref{fig:bd_voltage_calc}, which is consistent with the more diffusive nature of negative streamers \cite{Stollenwerk,zhu1996}.

\section{Summary and conclusions}
\indent 
The in-situ EFISH measurements were performed to investigate the decay of surface charges on alumina dielectrics after charging with an atmospheric pressure barrier discharge in N$_2$/C$_4$F$_7$N mixtures. 
A new method of the so called compensating voltage was developed for this aim. 
Additional method was developed using electrical measurements of breakdown voltage to assist the EFISH method which validated its results. 
Electrostatic simulations of the electric field distribution in the inter-electrode space together with the fluid modelling of the streamer development further support the result of the EFISH measurements, additionally being able to estimate the surface charge spread on the surface after its deposition. 
Furthermore, electrical current measurements enabled to follow the current pulses of the surface charging discharges with sub-nanosecond temporal resolution and few-milliamperes of amplitudes. 
Using these measurements, the amount of charge deposited during single discharge events was quantified, and this value served also as an input for the numerical model.
Moreover, the nonlinear hyperpolarizability tensor component of C$_4$F$_7$N was successfully determined.

The EFISH method using compensating voltage enabled to follow the surface charge decay for several days after charge deposition. The results show that admixture of C$_4$F$_7$N in nitrogen increases breakdown voltage as expected, yet surprisingly it prolongs the charge retention on the alumina surface. 
Using the isothermal surface charge decay model, charge traps in alumina dielectrics were identified - consistent with findings from experiments in nitrogen-containing atmospheres using conventional electrostatic probe techniques. This further supports the applicability of the EFISH method and highlights its future potential in this field.  
More specifically, measurements with C$_4$F$_7$N admixture indicate that the charge traps shift towards higher energy levels, which accounts for the slower charge decay.

The presented results further highlight the capability of EFISH-based diagnostics not only for in-situ monitoring of the discharge development phases, but also for tracking the evolution of residual surface charge afterward and studying of charge trapping on the dielectrics - establishing it as a powerful in-operando diagnostic tool for the entire discharge cycle investigation.

\vspace{1cm}

{\bf Acknowledgements}: This research was supported by the project Nr. TK04020069 funded by  the Technology Agency of the Czech Republic as a part of the Programme THE?TA, and by project CEPLANT (LM2023039) funded by the Ministry of Education, Youth and Sports of the Czech Republic. Portions of this research were carried out at the ELI Beamlines Facility, a European user facility operated by the Extreme Light Infrastructure ERIC.

\bibliographystyle{iopart-num}
\newcommand{\newblock}{}
\bibliography{ref}

\providecommand{\newblock}{}
\begin{thebibliography}{10}
\expandafter\ifx\csname url\endcsname\relax
  \def\url#1{{\tt #1}}\fi
\expandafter\ifx\csname urlprefix\endcsname\relax\def\urlprefix{URL }\fi
\providecommand{\eprint}[2][]{\url{#2}}

\bibitem{Stollenwerk}
Stollenwerk L, Laven J~G and Purwins H~G 2007 {\em Phys. Rev. Lett.\/} {\bf
  98}(25) 255001
  \urlprefix\url{https://link.aps.org/doi/10.1103/PhysRevLett.98.255001}

\bibitem{Leonov_2016}
Leonov S~B, Adamovich I~V and Soloviev V~R 2016 {\em Plasma Sources Science and
  Technology\/} {\bf 25} 063001
  \urlprefix\url{https://doi.org/10.1088/0963-0252/25/6/063001}

\bibitem{niem1995}
Niemeyer L 1995 {\em IEEE Transactions on Dielectrics and Electrical
  Insulation\/} {\bf 2} 510--528

\bibitem{bartnikas2002}
Bartnikas R 2002 {\em IEEE Transactions on Dielectrics and Electrical
  Insulation\/} {\bf 9} 763--808

\bibitem{rabie2018}
Rabie M and Franck C~M 2018 {\em Environmental Science \& Technology\/} {\bf
  52} 369--380 \urlprefix\url{https://doi.org/10.1021/acs.est.7b03465}

\bibitem{horky2024Holm}
Horký M, Kadlec S and Jimenez S~O 2024 Hot plasma decay and electrical
  breakdown in a miniature relay during hybrid switching {\em Proceedings of
  the 32nd International conference on electrical contacts and 69th IEEE Holm
  conference on electrical contacts\/} (Annapolis, MD, USA) p 209

\bibitem{vemul2023}
Vemulapalli H and Franck C~M 2023 {\em Journal of Physics D: Applied Physics\/}
  {\bf 56} 065202 \urlprefix\url{https://dx.doi.org/10.1088/1361-6463/acaab7}

\bibitem{liu2025}
Liu T, Sun H, Li G, Zhang Y, Wang J, Xiao J, Lu Y, Niu C and Wu Y 2024 {\em
  Journal of Physics D: Applied Physics\/} {\bf 58} 045203
  \urlprefix\url{https://dx.doi.org/10.1088/1361-6463/ad8bd7}

\bibitem{navarro2023}
Navarro-Rodriguez M, Palacios-Lidon E and Somoza A~M 2023 {\em Applied Surface
  Science\/} {\bf 610} 155437 ISSN 0169-4332
  \urlprefix\url{https://www.sciencedirect.com/science/article/pii/S0169433222029658}

\bibitem{Brandenburg}
Brandenburg R, Bogaczyk M, Höft H, Nemschokmichal S, Tschiersch R, Kettlitz M,
  Stollenwerk L, Hoder T, Wild R, Weltmann K~D, Meichsner J and Wagner H~E 2013
  {\em Journal of Physics D: Applied Physics\/} {\bf 46} 464015
  \urlprefix\url{https://dx.doi.org/10.1088/0022-3727/46/46/464015}

\bibitem{wild2014}
Wild R, Benduhn J and Stollenwerk L 2014 {\em Journal of Physics D: Applied
  Physics\/} {\bf 47} 435204
  \urlprefix\url{https://dx.doi.org/10.1088/0022-3727/47/43/435204}

\bibitem{tschiersch2014}
Tschiersch R, Bogaczyk M and Wagner H~E 2014 {\em Journal of Physics D: Applied
  Physics\/} {\bf 47} 365204
  \urlprefix\url{https://dx.doi.org/10.1088/0022-3727/47/36/365204}

\bibitem{Emelyanov_model}
Emelyanov O and Shemet M 2016 {\em Journal of Electrostatics\/} {\bf 81} 71--75
  ISSN 0304-3886
  \urlprefix\url{https://www.sciencedirect.com/science/article/pii/S0304388616300249}

\bibitem{svandova2024}
Svandova L, Pazderka M, Pribyl R, Stastny P, Kelar J, Kelar~Tucekova Z,
  Slavicek P, Trunec M and Cernak M 2024 {\em Journal of Physics D: Applied
  Physics\/} {\bf 57} 175201
  \urlprefix\url{https://dx.doi.org/10.1088/1361-6463/ad1f30}

\bibitem{zhang2018}
Zhang P, Zhou F, Zhang C, Yan Z, Li J, Jin H, Zhang H and Lu J 2018 {\em
  Ceramics International\/} {\bf 44} 12112--12117 ISSN 0272-8842
  \urlprefix\url{https://www.sciencedirect.com/science/article/pii/S0272884218308095}

\bibitem{kinder2008}
Kindersberger J and Lederle C 2008 {\em IEEE Transactions on Dielectrics and
  Electrical Insulation\/} {\bf 15} 941--948

\bibitem{molinie2024}
Molinié P 2024 {\em Journal of Electrostatics\/} {\bf 129} 103930 ISSN
  0304-3886
  \urlprefix\url{https://www.sciencedirect.com/science/article/pii/S0304388624000378}

\bibitem{mrkvi2023}
Mrkvičková M, Kuthanová L, Bílek P, Obrusník A, Navrátil Z, Dvořák P,
  Adamovich I, Šimek M and Hoder T 2023 {\em Plasma Sources Science and
  Technology\/} {\bf 32} 065009
  \urlprefix\url{https://dx.doi.org/10.1088/1361-6595/acd6de}

\bibitem{Kumada}
Kumada A, Okabe S and Hidaka K 2009 {\em Journal of Physics D: Applied
  Physics\/} {\bf 42} 095209
  \urlprefix\url{https://dx.doi.org/10.1088/0022-3727/42/9/095209}

\bibitem{kinder2008b}
Kindersberger J and Lederle C 2008 {\em IEEE Transactions on Dielectrics and
  Electrical Insulation\/} {\bf 15} 949--957

\bibitem{llovera2009}
Llovera P, Molinié P, Soria A and Quijano A 2009 {\em Journal of
  Electrostatics\/} {\bf 67} 457--461 ISSN 0304-3886 11th International
  Conference on Electrostatics
  \urlprefix\url{https://www.sciencedirect.com/science/article/pii/S0304388609000205}

\bibitem{zhu1996}
Zhu Y, Takada T, Sakai K and Tu D 1996 {\em Journal of Physics D: Applied
  Physics\/} {\bf 29} 2892
  \urlprefix\url{https://dx.doi.org/10.1088/0022-3727/29/11/024}

\bibitem{kumada2002}
Kumada A, Sugihara T, Chiba M and Hidaka K 2002 {\em Review of Scientific
  Instruments\/} {\bf 73} 1939--1944 ISSN 0034-6748 (\textit{Preprint}
  \eprint{https://pubs.aip.org/aip/rsi/article-pdf/73/4/1939/19035026/1939\_1\_online.pdf})
  \urlprefix\url{https://doi.org/10.1063/1.1458042}

\bibitem{li2022}
Li T, Yan H~J, Li J~Q, Schulze J, Yu S~Q, Song J and Zhang Q~Z 2022 {\em Plasma
  Sources Science and Technology\/} {\bf 31} 055016
  \urlprefix\url{https://dx.doi.org/10.1088/1361-6595/ac676e}

\bibitem{slikboer2018}
Slikboer E, Sobota A, Guaitella O and Garcia-Caurel E 2017 {\em Journal of
  Physics D: Applied Physics\/} {\bf 51} 025204
  \urlprefix\url{https://dx.doi.org/10.1088/1361-6463/aa9b17}

\bibitem{pribyl2024}
Pribyl R, Lexmaul J, Pazderka M, Stastny P and Kelar J 2024 {\em Plasma
  Chemistry and Plasma Processing\/}
  \urlprefix\url{https://doi.org/10.1007/s11090-024-10523-2}

\bibitem{limburg}
Limburg A, Franco R~M and Nijdam S 2024 Tracking electron and hole trap
  densities on commercial display glass {\em 2024 IEEE Conference on Electrical
  Insulation and Dielectric Phenomena (CEIDP)\/} pp 1--4

\bibitem{dogariu2017}
Dogariu A, Goldberg B~M, O'Byrne S and Miles R~B 2017 {\em Phys. Rev.
  Applied\/} {\bf 7}(2) 024024
  \urlprefix\url{https://link.aps.org/doi/10.1103/PhysRevApplied.7.024024}

\bibitem{chng2020electric}
Chng T~L, Starikovskaia S~M and Schanne-Klein M~C 2020 {\em Plasma Sources
  Science and Technology\/} {\bf 29} 125002

\bibitem{chng2022effect}
Chng T~L, Pai D~Z, Guaitella O, Starikovskaia S~M and Bourdon A 2022 {\em
  Plasma Sources Science and Technology\/} {\bf 31} 015010

\bibitem{nakamura2021}
Nakamura S, Sato M, Fujii T, Kumada A and Oishi Y 2021 {\em Phys. Rev. A\/}
  {\bf 104}(5) 053511
  \urlprefix\url{https://link.aps.org/doi/10.1103/PhysRevA.104.053511}

\bibitem{orr2020}
Orr K, Yang X, Gulko I and Adamovich I~V 2020 {\em Plasma Sources Science and
  Technology\/} {\bf 29} 125022
  \urlprefix\url{https://dx.doi.org/10.1088/1361-6595/aba989}

\bibitem{raskar2022}
Raskar S, Orr K, Adamovich I~V, Chng T~L and Starikovskaia S~M 2022 {\em Plasma
  Sources Science and Technology\/} {\bf 31} 085002
  \urlprefix\url{https://dx.doi.org/10.1088/1361-6595/ac8072}

\bibitem{billeau2024}
Billeau J~B, Cusson P, Dogariu A, Morozov A, Seletskiy D~V and Reuter S 2024
  {\em Appl. Opt.\/} {\bf 63} 5203--5207
  \urlprefix\url{https://opg.optica.org/ao/abstract.cfm?URI=ao-63-19-5203}

\bibitem{BNC}
Synek P, Zemánek M, Kudrle V and Hoder T 2018 {\em Plasma Sources Science and
  Technology\/} {\bf 27}

\bibitem{Sitz68}
Sitz P and Yaris R 1968 {\em The Journal of Chemical Physics\/} {\bf 49}
  3546--3557 ISSN 0021-9606 (\textit{Preprint}
  \eprint{https://pubs.aip.org/aip/jcp/article-pdf/49/8/3546/18858116/3546\_1\_online.pdf})
  \urlprefix\url{https://doi.org/10.1063/1.1670632}

\bibitem{Irving74}
Bigio I~J and Ward J~F 1974 {\em Phys. Rev. A\/} {\bf 9}(1) 35--39
  \urlprefix\url{https://link.aps.org/doi/10.1103/PhysRevA.9.35}

\bibitem{Nijdam_2020}
Nijdam S, Teunissen J and Ebert U 2020 {\em Plasma Sources Science and
  Technology\/} {\bf 29} 103001
  \urlprefix\url{https://doi.org/10.1088/1361-6595/abaa05}

\bibitem{Kulikovsky1997}
Kulikovsky A~A 1997 {\em Journal of Physics D: Applied Physics\/} {\bf 30}
  441–450 ISSN 1361-6463
  \urlprefix\url{http://dx.doi.org/10.1088/0022-3727/30/3/017}

\bibitem{Tungli2023}
Tungli J, Horký M, Kadlec S and Bonaventura Z 2023 {\em Plasma Sources Science
  and Technology\/} {\bf 32} 105009 ISSN 1361-6595
  \urlprefix\url{http://dx.doi.org/10.1088/1361-6595/acfcd8}

\bibitem{Crank_Nicolson_1947}
Crank J and Nicolson P 1947 {\em Mathematical Proceedings of the Cambridge
  Philosophical Society\/} {\bf 43} 50–67

\bibitem{vanLeer1974}
van Leer B 1974 {\em Journal of Computational Physics\/} {\bf 14} 361–370
  ISSN 0021-9991 \urlprefix\url{http://dx.doi.org/10.1016/0021-9991(74)90019-9}

\bibitem{lxcat1}
Carbone E, Graef W, Hagelaar G, Boer D, Hopkins M~M, Stephens J~C, Yee B~T,
  Pancheshnyi S, van Dijk J and Pitchford L 2021 {\em Atoms\/} {\bf 9} 16 ISSN
  2218-2004 \urlprefix\url{http://dx.doi.org/10.3390/atoms9010016}

\bibitem{lxcat2}
Pitchford L~C, Alves L~L, Bartschat K, Biagi S~F, Bordage M, Bray I, Brion C~E,
  Brunger M~J, Campbell L, Chachereau A, Chaudhury B, Christophorou L~G,
  Carbone E, Dyatko N~A, Franck C~M, Fursa D~V, Gangwar R~K, Guerra V,
  Haefliger P, Hagelaar G~J~M, Hoesl A, Itikawa Y, Kochetov I~V, McEachran R~P,
  Morgan W~L, Napartovich A~P, Puech V, Rabie M, Sharma L, Srivastava R,
  Stauffer A~D, Tennyson J, de~Urquijo J, van Dijk J, Viehland L~A, Zammit M~C,
  Zatsarinny O and Pancheshnyi S 2016 {\em Plasma Processes and Polymers\/}
  {\bf 14} ISSN 1612-8869
  \urlprefix\url{http://dx.doi.org/10.1002/ppap.201600098}

\bibitem{lxcat3}
Pancheshnyi S, Biagi S, Bordage M, Hagelaar G, Morgan W, Phelps A and Pitchford
  L 2012 {\em Chemical Physics\/} {\bf 398} 148–153 ISSN 0301-0104
  \urlprefix\url{http://dx.doi.org/10.1016/j.chemphys.2011.04.020}

\bibitem{lxcat}
TRINITI database, \url{www.lxcat.net/TRINITI}, retrieved on November 12, 2024

\bibitem{TRINITI}
Dyatko N, Kochetov I, Napartovich A and Sukharev A 2015 Eedf: The software
  package for calculations of electron energy distribution function
  \urlprefix\url{http://rgdoi.net/10.13140/RG.2.1.3675.1528}

\bibitem{bolsig}
Hagelaar G~J~M and Pitchford L~C 2005 {\em Plasma Sources Science and
  Technology\/} {\bf 14} 722–733 ISSN 1361-6595
  \urlprefix\url{http://dx.doi.org/10.1088/0963-0252/14/4/011}

\bibitem{Davies1971}
Davies A, Davies C and Evans C 1971 {\em Proceedings of the Institution of
  Electrical Engineers\/} {\bf 118} 816 ISSN 0020-3270
  \urlprefix\url{http://dx.doi.org/10.1049/piee.1971.0161}

\bibitem{Biagi}
Biagi database, \url{ www.lxcat.net/Biagi}, retrieved on November 12, 2024

\bibitem{Community}
Community database, \url{ www.lxcat.net/Community}, retrieved on November 12,
  2024

\bibitem{Shelton1990}
Shelton D 1990 {\em Physical Review A\/} {\bf 42} 2578

\bibitem{teyssedre2021}
Teyssedre G, Zheng F, Boudou L and Laurent C 2021 {\em Journal of Physics D:
  Applied Physics\/} {\bf 54} 263001
  \urlprefix\url{https://dx.doi.org/10.1088/1361-6463/abf44a}

\bibitem{simmons1973}
Simmons J~G and Tam M~C 1973 {\em Phys. Rev. B\/} {\bf 7}(8) 3706--3713
  \urlprefix\url{https://link.aps.org/doi/10.1103/PhysRevB.7.3706}

\bibitem{wang2019}
Wang F, Zhang T, Li J, Zeeshan K~M, He L, Huang Z and He Y 2019 {\em IEEE
  Transactions on Dielectrics and Electrical Insulation\/} {\bf 26} 702--737

\bibitem{ambrico2009}
Ambrico P~F, Ambrico M, Schiavulli L, Ligonzo T and Augelli V 2009 {\em Applied
  Physics Letters\/} {\bf 94} 051501 ISSN 0003-6951 (\textit{Preprint}
  \eprint{https://pubs.aip.org/aip/apl/article-pdf/doi/10.1063/1.3076122/14406690/051501\_1\_online.pdf})
  \urlprefix\url{https://doi.org/10.1063/1.3076122}

\bibitem{jansky2021}
J{\'{a}}nsk{\'{y}} J, Bessi{\'{e}}res D, Brandenburg R, Paillol J and Hoder T
  2021 {\em Plasma Sources Science and Technology\/} {\bf 30} 105008
  \urlprefix\url{https://doi.org/10.1088/1361-6595/ac2043}

\bibitem{papageorghiou2009}
Papageorghiou L, Panousis E, Loiseau J~F, Spyrou N and Held B 2009 {\em Journal
  of Physics D: Applied Physics\/} {\bf 42} 105201
  \urlprefix\url{https://dx.doi.org/10.1088/0022-3727/42/10/105201}

\bibitem{Viegas2022}
Viegas P, Slikboer E, Bonaventura Z, Guaitella O, Sobota A and Bourdon A 2022
  {\em Plasma Sources Science and Technology\/} {\bf 31} 053001 ISSN 1361-6595
  \urlprefix\url{http://dx.doi.org/10.1088/1361-6595/ac61a9}

\bibitem{Brzsnyi2008}
B\"{o}rzs\"{o}nyi A, Heiner Z, Kalashnikov M~P, Kovács A~P and Osvay K 2008
  {\em Applied Optics\/} {\bf 47} 4856 ISSN 1539-4522
  \urlprefix\url{http://dx.doi.org/10.1364/AO.47.004856}

\end{thebibliography}

\end{document}